# Direct-bonded diamond membranes for heterogeneous quantum and electronic technologies

Xinghan Guo[1], Mouzhe Xie[2,†], Anchita Addhya[1,†], Avery Linder[1,†], Uri Zvi [1],
Stella Wang[3], Xiaofei Yu[3], Tanvi D. Deshmukh[3], Yuzi Liu[4], Ian N. Hammock[1],
Zixi Li[1], Clayton T. DeVault[1,5], Amy Butcher[1], Aaron P. Esser-Kahn[1],
David D. Awschalom[1,3,5], Nazar Delegan[1,5], Peter C. Maurer[1,5],
F. Joseph Heremans[1,5], Alexander A. High[1,5,∗]

[1]*Pritzker School of Molecular Engineering, University of Chicago, Chicago, IL 60637, USA*
[2]*School of Molecular Sciences, Arizona State University, Tempe, AZ 85287, USA*
[3]*Department of Physics, University of Chicago, Chicago, IL 60637, USA*
[4]*Center for Nanoscale Materials, Argonne National Laboratory, Lemont, IL 60439, USA*
[5]*Center for Molecular Engineering and Materials Science Division,*
*Argonne National Laboratory, Lemont, IL 60439, USA*
† *These authors contributed equally to this work.*
∗*E-mail: ahigh@uchicago.edu*

**Diamond has superlative material properties for a broad range of quantum and electronic technologies. However, heteroepitaxial growth of single crystal diamond remains limited, impeding integration and evolution of diamond-based technologies. Here, we directly bond single-crystal diamond membranes to a wide variety of materials including silicon, fused silica, sapphire, thermal oxide, and lithium niobate. Our bonding process combines customized membrane synthesis, transfer, and dry surface functionalization, allowing for minimal contamination while providing pathways for near unity yield and scalability. We generate bonded crystalline membranes with thickness as low as 10 nm, sub-nm interfacial regions, and nanometer-scale thickness variability**

**over 200 by 200 μm² areas. We measure spin coherence times $T_2$ for nitrogen vacancy centers in bonded membranes of up to 623(21) μs, suitable for advanced quantum applications. We demonstrate multiple methods for integrating high quality factor nanophotonic cavities with the diamond heterostructures, highlighting the platform versatility in quantum photonic applications. Furthermore, we show that our ultra-thin diamond membranes are compatible with total internal reflection fluorescence (TIRF) microscopy, which enables interfacing coherent diamond quantum sensors with living cells while rejecting unwanted background luminescence. The processes demonstrated herein provide a full toolkit to synthesize heterogeneous diamond-based hybrid systems for quantum and electronic technologies.**

# Main

Diamond is broadly proposed for future quantum and electronic technologies. For example, color centers in diamond offer exceptional coherence properties[1,2] and robust spin-photon interfaces[3,4]. This enabled recent progress on quantum networking demonstrations[5,6] and quantum sensing applications, notably nuclear magnetic resonance (NMR) spectroscopy[7], magnetometry[8], and electrometry[9]. Further evolution of these technologies requires heterogeneous material platforms to expand on-chip functionalities, including nonlinear photonics, microfluidics, acoustics, electronics, detectors, and light sources. Moreover, diamond has best-in-class figures-of-merit for several applications in power electronics and these technologies can similarly benefit from heterogeneous integration[10–12]. However, single-crystal diamond-based heterostructures are challenging to be synthesized directly due to the technical difficulty of heteroepitaxial overgrowth[13]. Alternatively, multiple approaches have been demonstrated to integrate thin film diamond, utilizing Van der Waals forces[14,15] or intermediary bonding layers such as epoxy[16] and hydrogen silsesquioxane (HSQ)[17]. While promising, a truly generic approach—robust to fabrication and integration processes without introducing defects, decoherence sources, or superfluous materials—is still missing. In reference, multi-material heterostructures are commonly generated via wafer bonding, a manufacturing process that is foundational to modern electronic technologies. By chemically bonding films of disparate materials, wafer bonding allows high-quality crystalline materials to be combined in instances where direct growth processes are insufficient.

## Membrane Bonding

In this work, we introduce surface plasma activation based synthesis of diamond heterostructures where diamond membranes are directly bonded to technologically relevant materials, including silicon, fused silica, thermal oxide, sapphire, and lithium niobate ($LiNbO_3$), with the capability of pre-existing on-chip structures. The flow diagram of the full bonding process is shown in Fig.1. The fabrication process begins with membrane synthesis via smart-cut[18], followed by homoepitaxial diamond overgrowth and *ex situ* or *in situ* color center formation. Substrates are then patterned to define individual membrane shapes via either photo- or electron-beam lithography. Target membranes are undercut by selectively removing $sp^2$ carbon via electrochemical (EC) etching, leaving a small tether attached to the diamond substrate for deterministic manipulation. Membrane overgrowth, patterning, and EC etching are detailed in our previous work[19]. Here, we limit our membrane sizes to 200 µm by 200 µm squares. Larger and more intricate membrane shapes can be generated by extending the EC etching time and slightly modifying the patterning step of the process flow.

Following EC etching, we utilize templated area-controlled polydimethylsiloxane (PDMS)

stamps to transfer and manipulate membranes with improved process yield and scalability[20]. This transfer process is shown in Fig.1 (a). The PDMS stamps have two different patterns, allowing for smaller (PDMS1-stamp) and larger (PDMS2-stamp) contact areas, and by extension, adhesion strength (see section 1.2 in SI). PDMS1-stamp is used to break the diamond tether and pick up the membrane, whereas PDMS2-stamp is used for flipping the diamond membrane from PDMS1-stamp and subsequent placement. In both cases, the prominence of the adhesion region, which is 50 µm taller than the rest of the stamp, ensures only the targeted membrane is contacted. This method enables multiple membrane transfers following EC etching, which, in the future, can be automated into a single step for the entire diamond substrate.

Next, we remove the underlying diamond layer that was damaged by $He^+$ implantation. This improves the overall crystallographic quality and fully decouples the final membranes, which are isotopically purified with controlled doping, from the low-cost type-IIa diamond substrate. This thinning is performed via inductively coupled plasma (ICP) reactive ion etching (RIE). To protect the final bonded substrate from being etched, we thin the membrane by placing it on an intermediate fused silica carrier wafer. Intermediate wafers are coated with photo- (AZ1505) or electron beam resist (PMMA), which softens in the temperature range from 100 °C to 130 °C with reduced viscosity at subsequent stages. This additional step flips the membrane again so the growth side is facing up (exposed) on the target substrate, which eliminates growth side morphology constraints for bonding and enables precise depth control for near-surface and $\delta$-doped color centers. To prevent the resist from overheating and crosslinking, we developed a multi-cycle etching recipe with short plasma duration of ≤15 s per cycle. The schematic of the etched intermediate wafer is shown as the inset of Fig.1 (b); additional information can be found in section 1.4 of the SI. Using this methodology, we realize precise thickness control from 10 nm to 500 nm. The maximum thickness is determined by the homoepitaxial overgrowth step and can be modified to meet application needs.

We utilize a downstream $O_2$ plasma ashing for surface activation on both the diamond membrane and target substrate to enable subsequent bonding (Fig.1 (b)). The target substrates are subjected to a high power ashing recipe (gas flow 200 sccm, RF power 600 W for 150 s) with extended process duration for inert substrates such as sapphire and $LiNbO_3$. The membranes receive either this high power recipe or an $O_2$ descum clean (gas flow 100 sccm, RF power 200 W for 25 s), which does not etch nor roughen the diamond surface. The downstream $O_2$ plasma cleans and oxygen-terminates (see section 3.4 in SI) the membrane and carrier material surfaces without the need for wet processing.[21,22] To prevent functionalization degradation at elevated temperatures (see section 2.4 in SI), all ashing recipes are performed at room temperature.

Next, we bond the membrane to the target substrate, as shown in Fig.1 (c). We mount the patterned intermediate wafer onto a micropositioner-controlled glass slide via a flat, chip-size PDMS stamp. The target substrate is vacuum secured on a temperature-controlled stage. Leveraging optical access through the transparent intermediate wafer for alignment, we move the membrane to the target location and bring it into contact with the target substrate, which coincides with the appearance of membrane-scale interference fringes/patterns (see section 1.6 in SI). Using this method, we achieve an alignment precision of 30 μm and 0.1°. We sequentially heat the heterostructure by elevating the temperature of the stage through multiple steps (also see section 1.6 in SI). After reaching the resist softening point, we slide the intermediate wafer away, leaving the bonded structure behind. Future utilization of dedicated wafer-bonding equipment will significantly improve the precision and tolerance of all transfer steps.

Finally, to ensure a robust, covalently bonded interface between the membrane and the target wafer, we anneal the heterostructure at 550 °C under argon forming gas atmosphere to minimize undesired oxidation. This annealing also removes the polymethyl methacrylate (PMMA) residue and leaves a clean direct-bonded membrane as the final product if PMMA-based transfer is applied (see section 1.7 in SI for in-depth discussions).[23] A micrograph of a membrane-thermal oxide heterostructure with pre-defined markers is shown on the left of Fig.1 (d), revealing a high alignment accuracy. The right of Fig.1 (d) shows a membrane bonded to a fused silica trench, emphasizing our capability of bonding membranes to structured materials. The overall process yield stands above 95 %, limited by inconsistent plasma ashing chamber conditions and the poorly controlled approach angle of the transfer station, which can be readily improved by transitioning to process specific tooling.

**Material Characterization**

In-depth material characterization reveals the preserved diamond quality throughout the bonding process. We utilize atomic force microscopy (AFM) to characterize the membrane surface morphology, which is a critical determinant not only for successful plasma-activated bonding, but also the coherence and stability of near-surface color centers.[24] As shown in Fig.2 (a), both small and large area scanning results returned atomically flat surface profiles with $R_q \leq 0.3$ nm. Additionally, we characterize target substrates via AFM to ensure sub-nm roughness post plasma treatments as detailed in Table S1 of the SI.

Beyond local height variation, the bonded membrane shows a general flatness of $\approx 1$ nm, as characterized via profilometry. The thickness profile of the membrane, shown in Figure 2 (b), reveals a uniform height of 493.7 ± 1.1 nm, with the standard deviation less than the instrument resolution (1.5 nm) for large scale scanning. Beyond line scans, a two dimensional flatness map of the membrane is studied via confocal laser scanning microscopy, and is detailed in section 2.3 of the SI.

The effectiveness of the plasma surface activation is characterized by tracking the change in surface hydrophilicity of the bonding interfaces via contact angle measurements[25,26], as shown in Fig. 2 (c). For the diamond surface, the contact angle reduces from 52.4° ±0.7° to 6.1° post high power plasma treatment, indicating a considerable increase in hydrophilicity. This is confirmed via quantitative X-ray photoelectron spectroscopy (XPS) characterization of surface species also shown in Fig.2 (c) (see section 2.6 in SI for more details). We note that surface hydrophilicity is directly correlated with a reduction of surface amorphous carbon $sp^2$ bonds as quantified by the more reliable D-parameter extrapolation of the C KLL line[27,28]. Furthermore, the decrease from the raw C 1s quantification (likely an increase of ether-like terminations[24]) and an increase of surface available sapphire-O bonds indicate an effective surface preparation and oxygen termination to both surfaces. Similar behavior is confirmed on all target bonding materials with observed contact angles below 20° post treatments.

The quality of the bonded diamond heterostructure is directly studied via high resolution transmission electron microscopy (HRTEM). Fig.2 (d)-(e) shows an ICP-thinned (from ≈309 nm to $10 \pm 0.3$ nm while the lateral dimension remains to be 200 µm × 200 µm) diamond membrane bonded to a sapphire substrate. The thinness and uniformity reflect the high level of process control and allows single field of view characterization of both diamond membrane interfaces. The HRTEM image reveals several critical features. Firstly, the membrane exhibits uniform crystallinity and morphology throughout its thickness. Secondly, we observe a sharp, sub-0.5 nm interface between the crystalline diamond and sapphire. Thirdly, there is a repeating atomic arrangement throughout the interface profile, evidence of a covalently cross-linked interface.[29–32] Energy Dispersive X-ray Spectroscopy (EDS) analysis of the various elements associated with the intersection (C, Al, O) places an upper limit on the bonding interface to be less than 2 nm (see section 2.5 in SI). We note that the EDS analysis artificially broadens the interface as a result of the slight angular mismatch between the electron beam depth projection and the actual physical interface.

Furthermore, we characterize the optical properties of group IV color centers in the bonded membranes. Confocal imaging reveals that germanium vacancies (GeV$^-$) within the bonded membranes have high signal-to-background and sufficient optical coherence for applications in quantum technologies. Fig.2 (c) shows a typical photoluminescence (PL) map of individual GeV$^-$ centers hosted in a membrane bonded to a distributed Bragg reflector (DBR) mirror at 4 K. The signal-to-background ratio of GeV$^-$ can be as large as $\approx 65$ with an average value of $\approx 40$, a significant improvement from any suspended HSQ-based membranes (5 to 30) and bulk diamond ($\approx 25$).[19] The bonding process also preserves GeV$^-$ centers' optical coherence and introduces minimal strain, as shown in the section 3.2 and 3.3 of SI.

Lastly, to examine the compatibility of qubits in direct-bonded membranes with quantum technologies, we investigate the spin properties of NV$^-$ centers in an $\approx 150$ nm-thick diamond membrane bonded to a thermal oxide substrate. Room temperature measurements of a NV$^-$ center show a remarkably long Hahn-echo $T_2$ of 632(21) µs (Fig.2 (g)) and a Ramsey $T_2$ of 92(16) µs (SI). These coherence times are comparable to the 600 µs to 2000 µs coherence times reported in isotopically-purified bulk diamond[33] and are suitable for high-performance quantum sensing[34–36]. The spin echo oscillation originates from the nuclear spin of a nearby $^{13}$C atom (see section 5.7 of SI for details).

## Quantum technologies with bonded diamond membranes

Here we demonstrate the suitability of diamond-based heterogeneous material platforms for quantum technologies. First, we explore nanophotonic integration, which improves qubit addressability and is broadly utilized in quantum photonics. Photonic integration is commonly achieved by patterning diamond into undercut, suspended structures, creating geometrical constraints that complicate further multiplexing and integration with on-chip single-photon detectors, electronics, or other devices that could enhance quantum network functionality. Here, we show that our bonded membranes enable multiple approaches to photonic integration.

First, we utilize templated atomic layer deposition (ALD) of TiO$_2$ to create nanophotonic devices on the surface of a 50 nm-thick membrane bonded to a fused silica substrate (see section 4.2.1 in SI for fabrication details).[37] The schematic is shown as the upper image of Fig.3 (a), with excitation and collection ports (grating couplers) colored in red and blue, respectively. In these devices, the optical mode effectively hybridizes between the $TiO_2$ and diamond. Separate microscope images of $TiO_2$ fishbone cavities and ring resonators are shown in Fig.3 (b). These images were taken at the same location but different fabrication rounds, highlighting the robustness of the membranes to cleanroom processing and the recyclability of photonic integration.

Transmission measurements of the cavities reveal significant improvement of quality factors. The transmission spectrum of a typical fishbone cavity with target wavelength 737 nm — the wavelength of SiV emission — is shown in Fig.3 (c). The highest measured quality factor $Q$ is 10 640, with a three device average of 10150±350. These values are 2.5 times higher than our previous demonstration[37], resulting from elimination of the bonding layer and the improvement of the diamond crystal quality. We also predict a maximum Purcell enhancement factor of 270 in the diamond based on updated $Q$ factors. These metrics are suitable for state-of-the-art experiments in cavity quantum electrodynamics.[38]. Similarly, ring resonators measured through the drop port exhibit quality factors as $Q_{TM}$ = 16319 ($Q_{TE}$ = 12620) for transverse magnetic, TM (transverse electric, TE) modes (see section 4.2.2 in the SI for details).

Next, to explore the robustness of our bonded diamond platform to direct photonics integration, we etch nanophotonic ring resonators directly into a 280 nm-thick diamond membrane using a lithographically defined hard mask and a single RIE step. This is simplified in comparison to the angular etching or isotropic etching methods commonly used to create suspended diamond photonics from monolithic bulk diamond.[15,39] The schematic is shown in the lower image of Fig.3 (a). Bright and dark field images of the devices are shown in Figure 3 (d), demonstrating the high quality and uniformity of the fabrication process. As plotted in Fig.3 (e), these ring resonators exhibit quality factors $Q$ of 21 883 at visible wavelength range, with large field confinement within the diamond. Although this value looks slightly lower than the best-reported visible-wavelength quality factors $3 \times 10^4$ to $6 \times 10^4$ for diamond[40], our finesse $F = 188.4$ is higher than the value calculated from the previous work ($\approx 69.4$) due to the use of a much smaller ring diameter, and thus, a larger free spectral range (FSR). Combined with a separate demonstration of high quality photonic crystal cavities with $Q$ up to $1.8 \times 10^5$ in our direct-bonded membranes that are subsequently suspended[41], our platform supports a broad range of photonic integration. Additionally, our platform enables direct integration with other visible-frequency photonic platforms including lithium niobate, silicon nitride, and titanium dioxide, as well as on-chip light sources and detectors, and thus paves a path for hybrid quantum photonic technologies.

Diamond heterostructures also have distinct advantages in quantum metrology, such as nanoscale magnetic field[42,43], electric field[44], and temperature[45,46] sensing. In these applications a diamond sensor is typically brought in close proximity of a sensing target. The most sensitive diamond sensors rely on high-purity single-crystals with the sensing target bound on the diamond's top surface[47]. The large thickness and refractive index of conventional bulk diamond requires optical initialization and readout of the sensing qubits from the top surface. As a consequence of this geometry, the target systems need to possess optical transparency, low auto-fluorescence, and high photo-stability, which are significant restrictions for the study of biological systems. Bonded membranes overcome these challenges by enabling optical addressability through the back of the membrane, without the need for passing through the top surface and the sensing target.

First, we investigate the stability and addressability of individual nitrogen vacancy (NV$^-$) centers (i.e., our qubit sensor) in bonded diamond membranes. Figure 4 (a)-(c) shows individually resolvable photostable NV$^-$ centers in a 160 nm-thick membrane bonded to a fused silica coverslip in widefield and confocal imaging. These emitters are confirmed to be NV$^-$ centers through the presence of their characteristic 2.87 GHz zero-field splitting by optically detected magnetic resonance (ODMR) spectroscopy (Fig. 4 (d)). NV⁻ centers in membranes were previously reported to have excellent spin coherence.[19] Next, we chemically functionalized the heterostructure surface using a technique recently developed for bulk diamond[48]. Using back illumination of our diamond membranes, we show that the fluorescence of individual Alexa488-labeled streptavidin molecules (Fig. 4 (e)) and streptavidin-conjugated Qdot-525 quantum dots (Fig. 4 (f)) can be detected. This enables localization of not only NV- centers but also the fluorescent sensing targets bound to the membrane's surface (also see section 5.3.4 in the SI for simultaneous detection). The ability to fluorescently detect the position of individual NV⁻ centers and proteins is important for NV-based single-molecule nuclear magnetic resonance[49] and electron paramagnetic resonance[50] spectroscopy, as this will allow for the efficient identification of NV⁻ centers that have a desired molecular target within the sensing range.

We also combine our bonded diamond membranes with TIRF-microscopy to demonstrate imaging at a reduced level of background luminescence. Fig. 4 (g) shows a schematic representation of a diamond membrane integrated in a flow channel with macrophage-like RAW cells grown on the diamond surface. Optical excitation above the critical angle ensures that only a small section above the diamond membrane is excited by the optical field. Staining the toll-like receptor 2 (TLR2) with Alexa488-labeled anti-TLR2 antibody reveals in TIRF imaging the location of individual protein distributed across cell surface (Fig. 4 (i)). This is in stark contrast with the epiluminescence mode where background luminescence prevents the imaging of individual molecules (Fig. 4 (h)). We note that the larger index of refraction ($n = 2.4$) of diamond results in an evanescent field that falls 1.6-times faster off compared with a conventional glass microscope coverslip ($n = 1.5$) (see section 5.6 in SI). Likewise, we show the sedimentation of living *Escherichia coli* bacteria on the diamond membrane introduced via a flow channel (see SI Video S1). Experiments enabled by the flow channel demonstrate remarkable flexibility to interface target samples with quantum diamond sensors, which is challenging to be fulfilled through conventional approaches.[51]

## Conclusions

We have demonstrated a complete process flow to create diamond-based heterogenous materials and technologies. The bonded membranes combine isotopic engineering, in-situ doping, and precise thickness control, while maintaining the surface morphology, flatness, and crystal quality necessary for quantum technologies. We generated bonded, continuous crystalline films as thin as 10 nm, well below previous demonstrations and comparable to material geometries in state-of-the-art microelectronics.[17,19,52] HRTEM reveals ordered, sub-nanometer bond interfaces, PL measurements demonstrate high signal-to-background ratio for all hosted color centers, and nitrogen vacancy centers maintain bulk-like spin coherence. The process is compatible with nano-structured substrates, has a compact footprint and requires no post-bond etching, ensuring the integrity of pre-existing target substrate structures. Bonded membranes are robust to multiple subsequent nanofabrication steps, and our method is compatible with standard semiconductor manufacturing processes including wafer-bonding.

Crucially, by avoiding intermediary adhesion materials, we generate optimal material heterostructures for applications in quantum photonics and quantum biosensing. Technological suitability for quantum photonics is demonstrated via the integration of high quality factor nanophotonics by either $TiO_2$ deposition or direct diamond patterning and etching. These diamond-based heterostructures, with minimal optical loss, are ideal candidates for on-chip nanophotonic integration and spin-photon coupling devices. Furthermore, we demonstrate that diamond membrane bonding unlocks novel experimental possibilities for quantum biosensing and imaging by integrating flow channels with diamond membranes. The simultaneous resolution of fluorescent molecules and $NV^-$ centers will enable accurate identification of proximal $NV^-$ sensors for desired sensing targets. The ultrathin diamond membranes also allow for TIRF illumination which strongly improves the signal contrast of local sensing targets while minimizing undesired laser excitation.

Our manufacturing process opens up a broad range of heterogeneous diamond-based platforms for quantum technologies. The integration of diamond with electro-optical and piezoelectric materials such as $LiNbO_3$ will pave the ways for on-chip, electrically-reconfigurable nonlinear quantum photonics and allow studies of quantum spin-phonon interactions[53,54]. Diamond bonding unlocks additional coupling possibilities with other solid state qubits, magnonic hybrid systems, or superconducting platforms[35,55–57]. Furthermore, combining our diamond membranes with established techniques for the creation of highly coherent near-surface NV⁻ centers[19,24] will result in ultra-sensitive diamond-probes optimized for the study of molecular binding assays[58], two dimensional dichalcogenides (TMD)[59], and thin-film magnetic materials[60]. Lastly, with high thermal conductivity, large bandgap and high critical electric field, bonded diamond membranes have myriad applications in high power electronics[10–12].

# Acknowledgements

This work is primarily funded through Q-NEXT, supported by the U.S. Department of Energy, Office of Science, National Quantum Information Science Research Centers. Growth related efforts were supported by the U.S. Department of Energy, Office of Basic Energy Sciences, Materials Science and Engineering Division (N.D.). The membrane bonding work is supported by NSF award AM-2240399. The quantum metrology and sensing demonstration is supported by the U.S. National Science Foundation (NSF) Quantum Idea Incubator for Transformational Advances in Quantum Systems (QII-TAQS) for Quantum Metrological Platform for Single-Molecule Bio-Sensing (NSF OMA-1936118), and Quantum Leap Challenge Institute (QLCI-CI) for Quantum Sensing in Biophysics and Bioengineering (QuBBE)(NSF OMA-2121044). This work made use of the Pritzker Nanofabrication Facility (Soft and Hybrid Nanotechnol-ogy Experimental Resource, NSF ECCS-2025633) and the Materials Research Science and Engineering Center (NSF DMR-2011854) at the University of Chicago. Work performed at the Center for Nanoscale Materials, a U.S. Department of Energy Office of Science User Facility, was supported by the U.S. DOE, Office of Basic Energy Sciences, under Contract No. DE-AC02-06CH11357. The quantum photonics work and part of the membrane bonding work is supported by the Quantum Leap Challenge Institute for Hybrid Quantum Architectures and Networks (HQAN) (NSF OMA-2016136). A. Addhya additionally acknowledges support from Kadanoff-Rice fellowship (NSF DMR-2011854). C. DeVault receives support from the CQE IBM postdoctoral fellowship training program. The authors thank Dr. David A. Czaplewski for valuable discussion and experimental guidance, Dr. Kazuhiro Kuruma, Alexander Stramma, Hope Lee and Zander Galluppi for providing materials, and Dr. Peter Duda and Dr. Yizhong Huang for nanofabrication assistance.

## Competing interest

Guo X, High A, Deshmukh T, Linder A, Hammock I, Delegan N, Devault C, and Heremans J. filed a PCT (International) patent for the methods of bonding diamond membranes.

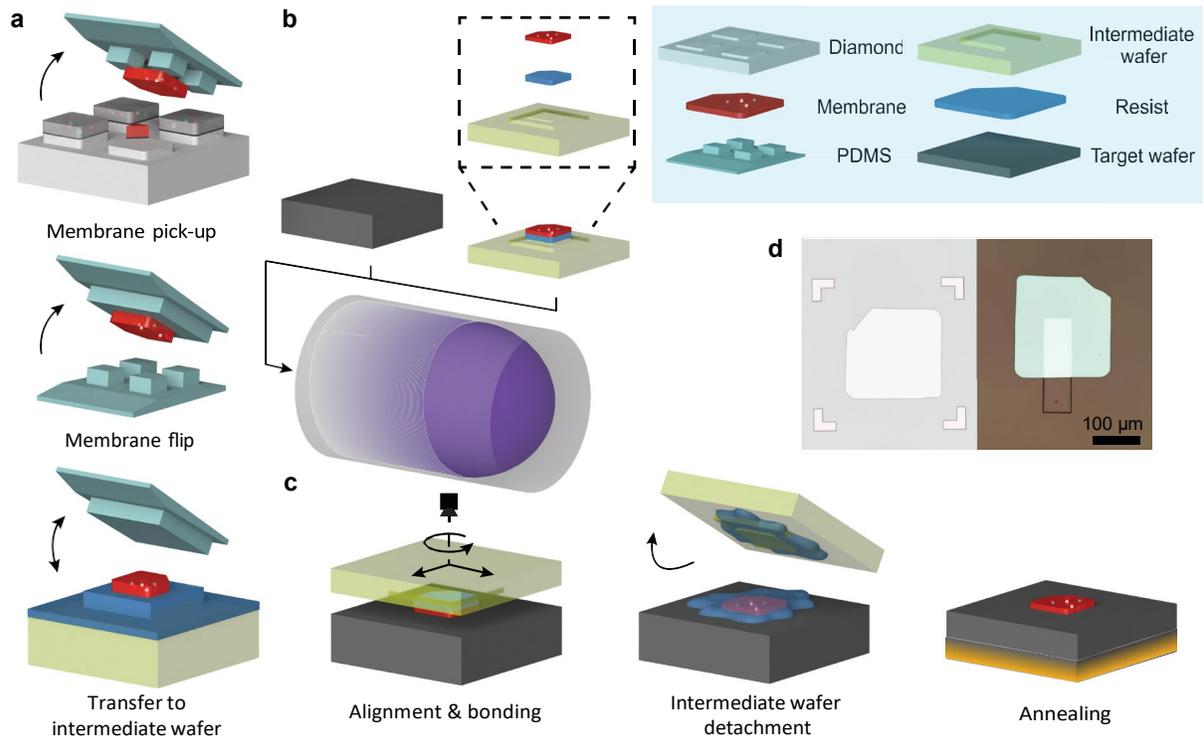

Figure 1: Schematics of the plasma-activated bonding of diamond membranes. (a) Diamond membrane transfer to the intermediate wafer. From top to down: membrane pick-up from the diamond substrate using PDMS1-stamp, membrane flipping with PDMS2-stamp, membrane placement to a photoresist or electron beam resist covered intermediate wafer. (b) Diamond back etching and downstream oxygen plasma treatment. Inset: the detailed layer stack of the ICP-etched intermediate wafer. (c) Plasma-activated membrane bonding. Left to right: membrane alignment and bonding, temperature-controlled intermediate wafer detachment, and post-bonding annealing. (d) Microscope images of 155 nm-thick diamond membranes bonded to a thermal oxide substrate with markers (left) and a fused silica substrate with a 5 μm-deep trench etched prior to bonding (right).

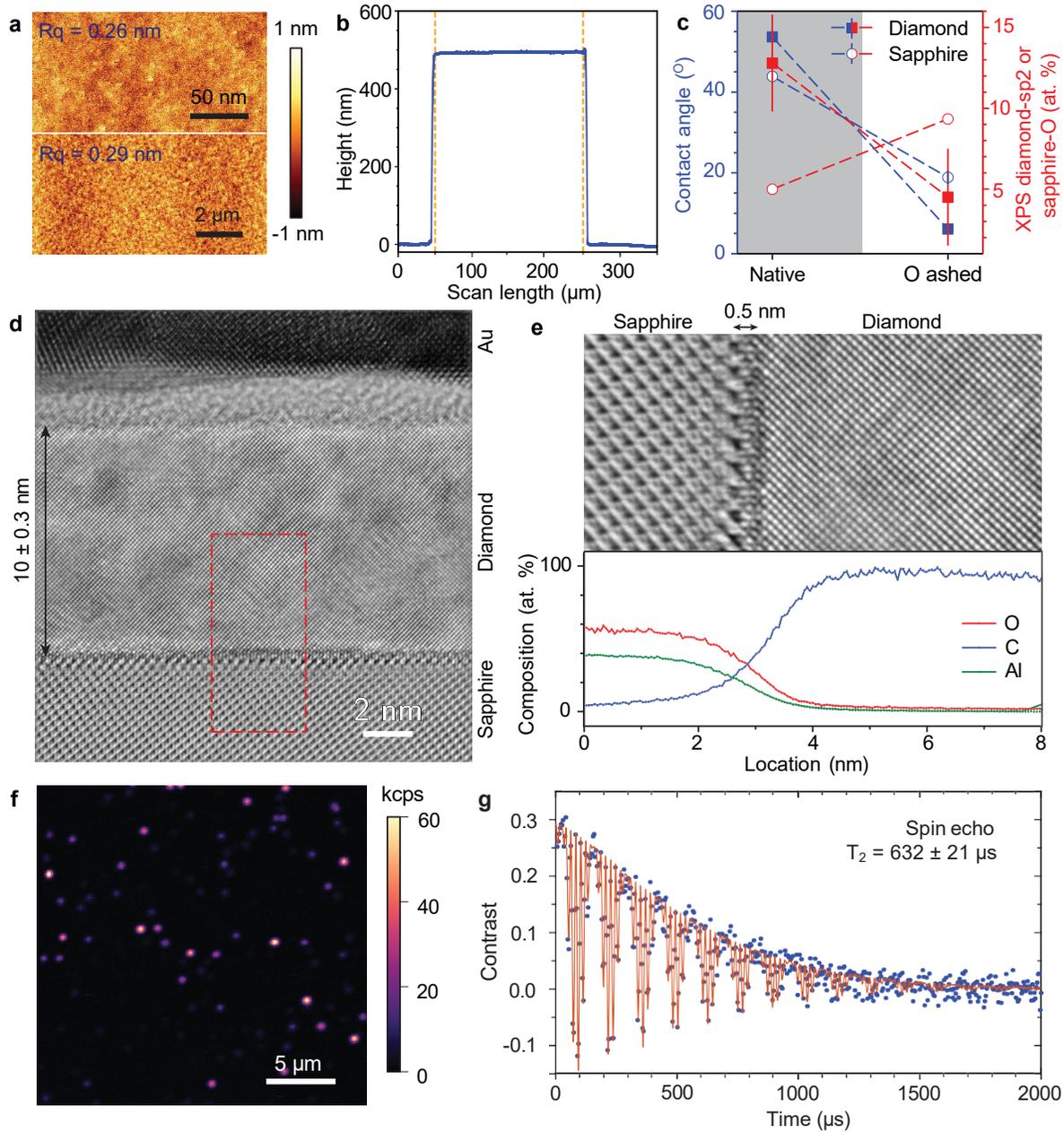

Figure 2: Characterization of the bonded membrane. (a) AFM of the diamond bonding interface (the etched side) post ICP etching. Atomically flat surfaces with $R_q \leq 0.3$ nm were observed in both small (200 nm by 100 nm, the upper figure) and large (10 μm by 5 μm, the lower figure) scanning areas. (b) Profilometry of a membrane-silicon heterostructure. The membrane region is highlighted by two dashed orange lines. The thickness of the membrane is 493.7 nm with a standard deviation of 1.1 nm. (c) The contact angle and XPS of diamond and sapphire pre- and post- high power plasma treatments. An increase of hydrophilicity is observed via the decrease of the contact angle, and the effect of oxygen termination is observed through the reduction of the carbon $sp^2$ as obtained from C KLL extrapolation of the $sp^2/sp^3$ ratio and the enhancement of the sapphire-O signals as obtained from the O 1s peak quantification. (d) HRTEM image of a 10 nm-thick membrane bonded to a c-plane sapphire substrate. The 2 nm intermediate layer on top of diamond comes from the lack of surface control before gold deposition. (e) Top: the zoomed-in HRTEM image of the diamond-sapphire bonding interface, the red dashed rectangle region in (e), showing a sub-0.5 nm thickness of the bonding intersection. Bottom: EDS elemental analysis across the bonding interface. (f) The PL map of GeV centers in a membrane bonded to a DBR mirror at 4 K. The signal-to-background ratio around the zero phonon line (ZPL) can be as high as 65, with the signal surpassing 65 kc s$^{-1}$. (g) Hahn-echo measurements of one typical NV$^-$ at room temperature showing a $T_2$ value of 632(21) μs. See section 5.7 of SI for data acquisition and fitting details.

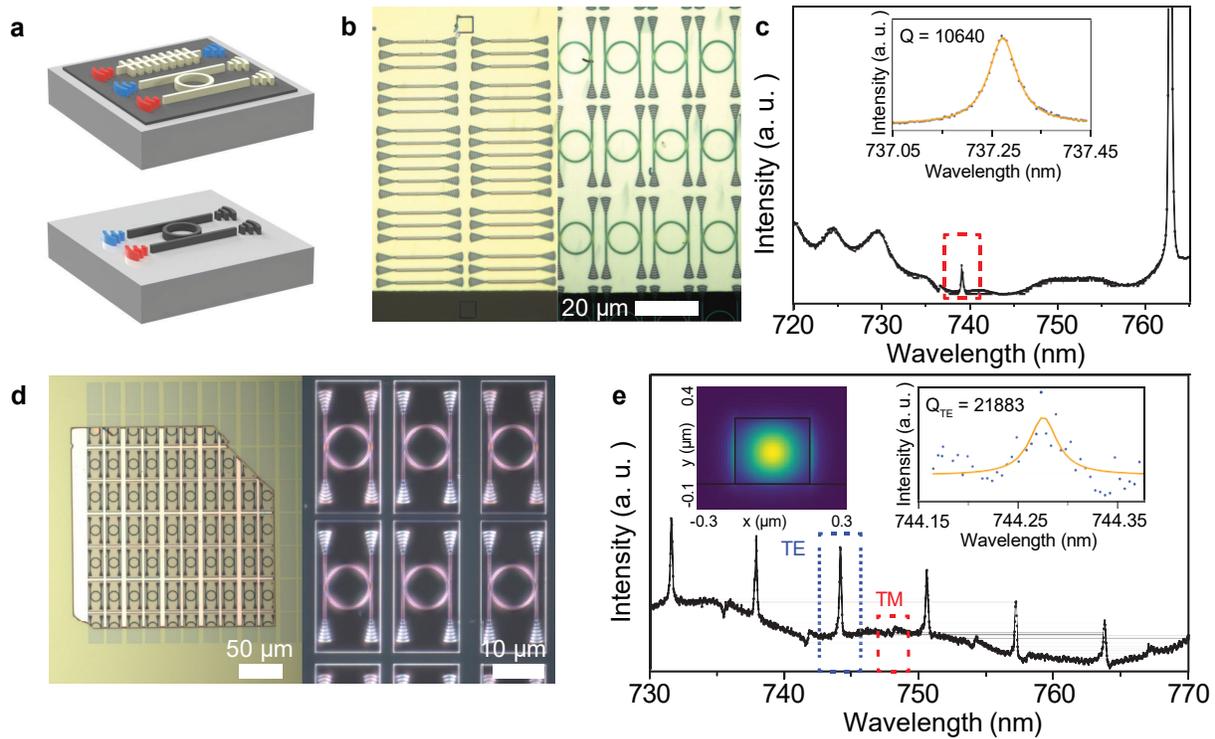

Figure 3: Nanophotonic integration with direct-bonded membranes. (a) Schematics of TiO$_2$-based (top) and diamond-based (bottom) nanophotonics on diamond membrane heterostructures. In this work fused silica (thermal oxide silicon) wafers are used as carrier wafers for the TiO$_2$ (diamond)-based demonstrations. The grating couplers for excitation (collection) are colored in red (blue), respectively. (b) Microscope images of TiO$_2$ fishbone cavities and ring resonators on a 50 nm-thick diamond membrane. Images were taken at the same location but different fabrication rounds. (c) The transmission spectrum of a fishbone cavity with resonant frequency at 737.26 nm. Inset: the transmission of the cavity with a tunable laser as the excitation source, showing a quality factor $Q$ of 10640 ± 118. (d) The bright field and dark field microscope images of the ICP-etched diamond ring resonators on a thermal oxide silicon substrate, showing great uniformity with minimal process contamination. (e) The transmission spectrum of the diamond-based ring resonator measured at the drop port. Insets: the TE mode profile, and the TE cavity resonance with a quality factor $Q_{TE}$ of 21883 ± 6284. The fluctuation in the right inset is caused by the instability of the optical setup.

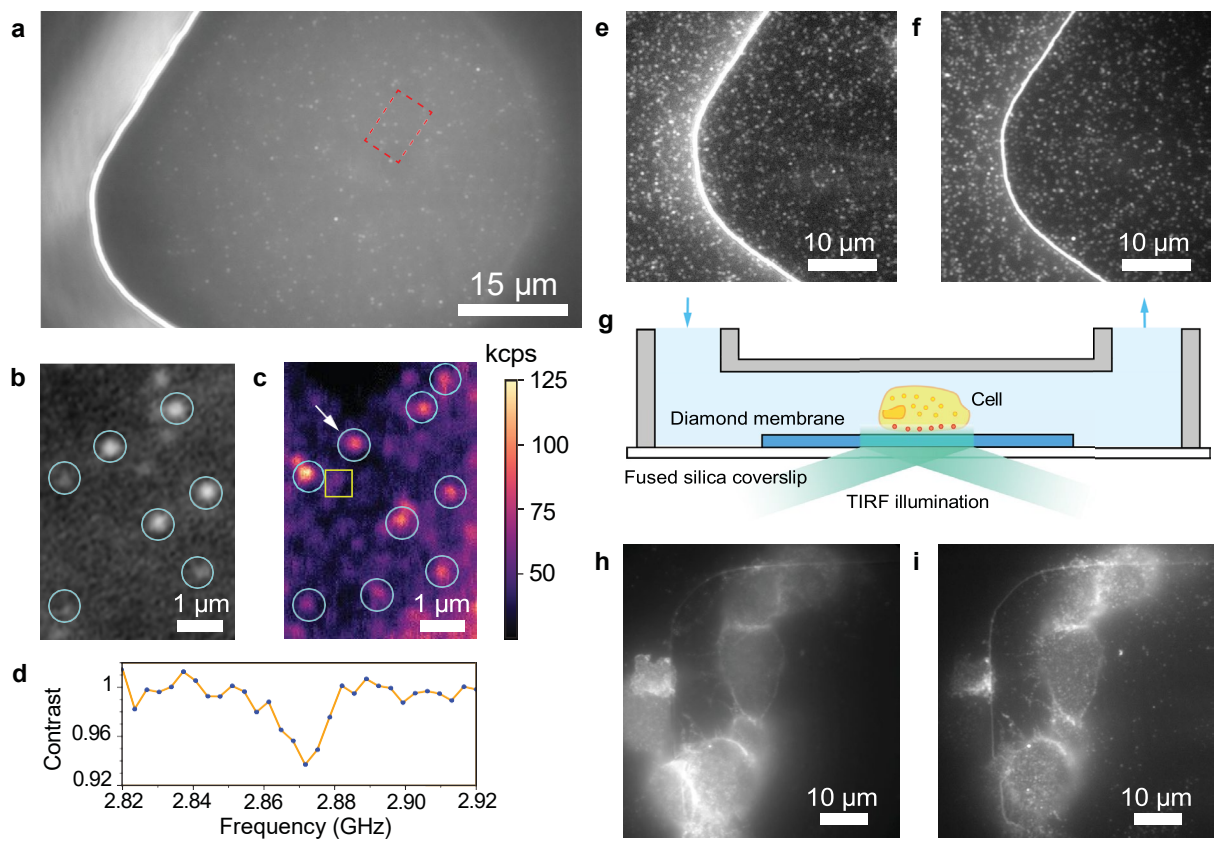

Figure 4: Imaging of NV$^-$ centers and surface-attached target molecules and cells in a flow channel. (a) Widefield fluorescence microscopy image of a diamond membrane corner containing NV$^-$ centers at room temperature. Only a round area in the center was illuminated to avoid back-reflection from membrane edges. (b) The zoomed-in image of the boxed region shown in (a) post rotation. (c) A confocal scan of the same region as (b) using a separate setup at room temperature. Emitters confirmed to be (not) NV$^-$ centers are highlighted in cyan circles (yellow boxes). Same symbols are used in (b). (d) A representative CW-ODMR spectrum from the NV$^-$ center labelled with a white arrow in (c). Additional studies of NV$^-$ spin coherence are included in the section 5.7 of SI. (e-f) Widefield fluorescence microscopy images of (e) Alexa-488-labeled streptavidin protein and (f) streptavidin-conjugated Qdot-525 quantum dots that were immobilized at the same region shown in (a) via biotinylated surface functionalization. (g) Schematic illustration of the flow channel structure and a cell illuminated by total internal reflection through the diamond membrane. (h-i) Fluorescence microscopy images of Alexa-488-labeled TLR2 receptors on RAW cell surfaces, under (h) episcopic and (i) objective-based TIRF illumination. Edges of the diamond membrane are also visible in these images.

# Supplementary information: Direct-bonded diamond membranes for heterogeneous quantum and electronic technologies

Xinghan Guo[1], Mouzhe Xie[2,†], Anchita Addhya[1,†], Avery Linder[1,†], Uri Zvi [1], Stella Wang[3], Xiaofei Yu[3], Tanvi D. Deshmukh[3], Yuzi Liu[4], Ian N. Hammock[1], Zixi Li[1], Clayton T. DeVault[1,5], Amy Butcher[1], Aaron P. Esser-Kahn[1], David D. Awschalom[1,3,5], Nazar Delegan[1,5], Peter C. Maurer[1,5], F. Joseph Heremans[1,5], Alexander A. High[1,5,∗]

[1]*Pritzker School of Molecular Engineering, University of Chicago, Chicago, IL 60637, USA*
[2]*School of Molecular Sciences, Arizona State University, Tempe, AZ 85287, USA*
[3]*Department of Physics, University of Chicago, Chicago, IL 60637, USA*
[4]*Center for Nanoscale Materials, Argonne National Laboratory, Lemont, IL 60439, USA*
[5]*Center for Molecular Engineering and Materials Science Division,*
*Argonne National Laboratory, Lemont, IL 60439, USA*
† *These authors contributed equally to this work.*
∗*E-mail: ahigh@uchicago.edu*

# 1 Fabrication process of membrane heterostructures

## 1.1 Diamond membrane synthesis and patterning

The diamond membrane synthesis follows the same method as described in our previous work.[1] The membrane tether layer was created via He$^+$ implantation (dose $5 \times 10^{16}$ cm$^{-2}$, energy 150 keV), followed by multi-step annealing (400 °C for 8 h, 800 °C for 8 h, and 1200 °C for 2 h) in a forming gas (4 % H$_2$, 96 % Ar) environment. Membrane overgrowth was performed in a microwave plasma chemical vapor deposition (MPCVD) chamber at Argonne National Laboratory. Four membrane substrates discussed in this work have overgrown layers of 185 nm, 260 nm, 400 nm and 660 nm, respectively. Overgrown substrates received either ion implantation (Si$^+$, Ge$^+$, Sn$^+$ or N$^+$) or $\delta$-doping $^{15}$N for color center creation. Substrates were then patterned and inductively coupled plasma (ICP)-etched to individual 200 µm by 200 µm membranes. In preparation of the pick-up step, membranes were undercut from substrates via electrochemical (EC) etching in fresh de-ionized water, leaving the membranes solely attached to the substrates by a small, breakable tether.

## 1.2 Membrane pick-up and transfer via patterned PDMS stamps

PDMS stamps were prepared by applying freshly mixed PDMS base and curing agent (Sylgard 184, Dow Corning) to a 4 inch silicon wafer with lithographically defined SU-8 (3050, ≈55 µm thick post baking) structures. Over 200 PDMS stamps with two different shapes can be generated on a single wafer. The PDMS1-stamp consists of four little square-shaped contacting fingers with 70 µm spacing, in order to break the diamond tether and pick up the membrane. The PDMS2-stamp contains a single large square with 300 µm length to flip and place the membrane. The membrane pick-up, flipping, and placement were carried out on a probe station (Signatone S1160). Schematics and microscope images of this process are shown in Figure S1 (a)-(d). In this work, we placed our membrane onto a resist-coated, pre-patterned fused silica substrate referred as intermediate wafer, as discussed in section 1.3. Beyond this work, the patterned PDMS stamps can also be applied to HSQ-based membrane placements[1] to improve the device yield.

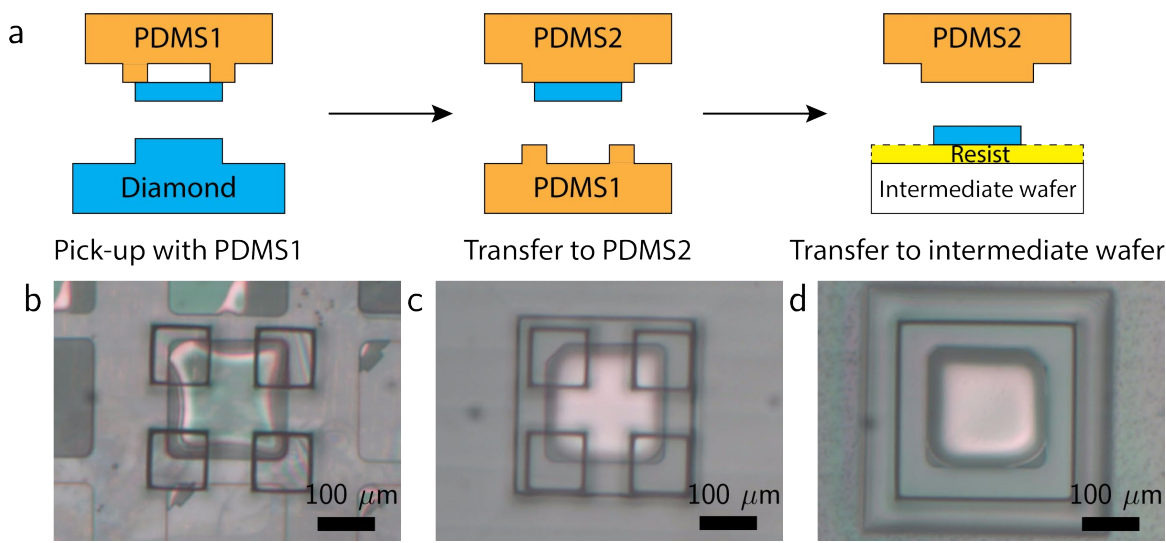

Figure S1: Deterministic diamond membrane transfer via patterned PDMS stamps. (a) Schematics of the membrane transfer with PDMS1-stamp and PDMS2-stamp. (b-d) Microscope images of (b) alignment and pick-up of the diamond membrane using PDMS1-stamp, (c) the membrane flipping using PDMS2-stamp, (d) membrane placement onto the intermediate wafer coated with resist.

## 1.3 Intermediate wafer preparation

Intermediate wafers we used are 13 mm by 13 mm double side polished fused silica substrates. In principle, any polished transparent substrate can serve as intermediate wafers. Prior to dicing, the wafer was lithographically patterned and ICP-etched to form 5 µm tall pillars at the center of each chip. The size of the pillar is 400 µm by 400 µm, shown as the largest square in Figure S1 (d). These elevated pillars are critical to compensate for the weakly defined approaching angle of the micropositioner on the transfer station. Pillars can also help protect existed structures on the final wafer.

Prior to the membrane transfer, intermediate wafers were spin-coated with a thin layer of positive photoresist (AZ 1505, Microchemicals GmbH, ≈500 nm) or electron beam resist (950 K PMMA A4, MicroChem, ≈250 nm). Positive resists are necessary in this process due to their much reduced viscosity after reaching the softening temperature, which grants smooth detachment in heated environments post plasma-activated bonding. When the membrane was flipped and placed on a pillar, it was released from the PDMS2-stamp due to stronger adhesion of the resist. The PDMS2-stamp-covered surface (the damaged side) was exposed again for the subsequent etching step.

## 1.4 Damaged layer removal via multi-cycle ICP etching

As will be discussed in section 2.1, out-of-plane strain resulted from $He^+$ implantation barricades effective bonding. The damaged layer also introduces considerable fluorescence background, which is not desirable for most optical measurements. We address these issues by removing the damaged layer via ICP reactive ion etching (RIE) (Plasma-Therm ICP Chlorine Etch). This step also defines the thickness of the membrane. We note that this intermediate wafer approach is generic and can be applied to HSQ-based membrane transfer as well to protect the final wafer from being etched.

We developed a multi-cycle $Ar/Cl_2$-$O_2/Cl_2$-$O_2$ etching sequence, limiting the effective etching time to 15 s per cycle. Cycles are separated by 3 pump-and-purge sequences to remove the gaseous residue, maintaining a consistent chamber environment and thus a constant etching rate. Compared to continuous etching, this multi-cycle procedure also prevents the resist from softening by keeping the sample temperature low.

## 1.5 Plasma treatment on diamond bonding interface

We use a downstream plasma asher (YES-CV200 RFS Plasma Strip/Descum System, Yield Engineering Systems Inc.) to activate bonding interfaces. Although other gas options ($N_2$, Ar, etc.) are possible[2], we choose $O_2$-based plasma for the surface functionalization due to their better performance on diamond and other materials[3,4]. In this work, we use two recipes on diamond membranes and carrier wafers, referred as the $O_2$ descum and the high power recipe. Both recipes are run at room temperature, since a degradation of hydrophilicity is observed after a brief baking (90 °C on a hot plate for 1 min) post plasma treatments, as discussed in section 2.4.

We compare the two recipes according to three metrics measured on the resulting products, namely, surface morphology, surface hydrophilicity, and the optical performance of $GeV^-$ and $NV^-$ centers, and assessed the effects on both carrier wafer and diamond membrane surfaces. Results are discussed in section 2.2, 2.4, 3.2 and 3.4. For the carrier wafer choices, we tested fused silica and thermal oxide silicon. Both recipes lead to enhanced surface hydrophilicity while maintaining good surface morphology, which are favorable for the subsequent bonding process. We expand the high power recipe to all carrier substrates used in this work due to the better hydrophilicity state it prepares. For diamond membranes, while both recipes could improve the signal to background ratio for optical characterizations of color centers, we did observe increased number of particle-like contaminates under atomic force microscopy (AFM) on the treated surfaces post high power recipes. We therefore employ the $O_2$ descum recipe for diamond membrane as a standard process except for $NV^-$ sensing applications which requires a more adequate charge state preparation. The contamination can be eliminated by switching to more specialized tooling.

We note that the surface activation on the target substrate alone is effective for the bonding process, with the bonding quality as good as the ones with $O_2$ descum treatment. However, the lack of membrane treatment could result in a higher optical background, as discussed in section 3.2. Nonetheless, it provides an option for diamond membrane systems that are, for example, sensitive to $O_2$ plasma.

## 1.6   Direct bonding of the diamond membrane

The schematics of the bonding process is shown in Figure S2 (a). To start, the intermediate wafer is mounted on a glass cantilever controlled by a micropositioner (Signatone CAP - 946) using a flat, chip size PDMS stamp. The target substrate is held to a temperature-controlled stage by vacuum. Due to the lack of full tilt angle control of the micropositioner, we could only set the approaching angle to 0° along one direction, leaving the other to be a small but weakly defined value. A bright-field camera allows us to monitor the diamond membrane through the transparent PDMS and intermediate wafer and align it to the desired location on the target substrate. The alignment precision is 30 µm, and the in-plane angle precision is 0.1°, both limited by the micropositioner tilt angle used in this work. Post alignment, we slowly bring down the intermediate wafer until part of the membrane is in contact with the target substrate, indicated by an appearance of interference pattern as exemplified in Figure S2 (b). We then step-wise increase the temperature (75 °C, 95 °C, and 125 °C for AZ 1505, 90 °C, 130 °C, and 170 °C for PMMA) , allowing the resist to reach thermal equilibrium at each stage. Abrupt temperature changes can cause undesirable resist re-flow across regions and impact the bonding quality. Once the temperature reaches the highest stage, the resist layer thoroughly softens, and the intermediate wafer tends to shift translationally to release stress, as shown in Figure S2 (c). Next, the intermediate wafer is slowly moved away from the membrane and lifted via the motorized stage, leaving the membrane on the target substrate covered by some residual resist, as shown in Figure S2 (d). Finally, the bonded heterostructure is left to cool down till room temperature, preparing for the subsequent annealing. Stripping resist prior to the annealing is not recommended because of the weak bonding quality at this point.

## 1.7   Membrane annealing and resist removal

The quality of the plasma-activated bonding highly depends on the formation of the covalent bonds, which can be greatly strengthened by additional annealing. The annealing can also remove the −OH bonds inside the bonding interface if present[3,4]. We anneal the heterostructure at 550 °C for 8 h to 14 h in argon forming gas environment (96 % of Ar, 4 % of $H_2$, ≈1 atm), which eliminates potential oxidization on the diamond surface. Bonding strength when annealed at 450 °C is not sufficient as the membrane can detach from the carrier substrate during acid cleaning. This might be explained by a less inert diamond surface to oxygen when the temperature exceeds 450 °C, which is about the temperature for standard diamond oxygen termination[5].

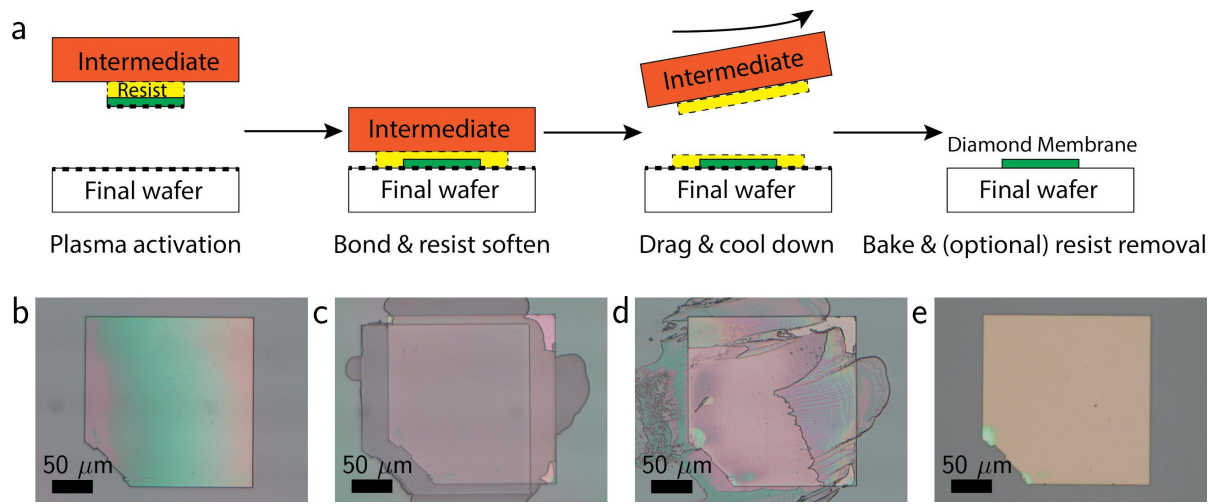

Figure S2: Plasma-activated membrane bonding. (a) Schematics of the bonding process. (b)-(e) Microscope images of (b) membrane alignment and initial contact, with the interference pattern induced by a non-zero approaching angle, (c) membrane bonded to the target wafer with resist re-flow at elevated temperatures, (d) the bonded membrane with residual resist after lifting the intermediate wafer, (e) the membrane post baking and optional resist removal. Here the AZ 1505 photoresist was applied which requires a di-acid clean, but PMMA-based bonded membrane is clean post annealing.

If PMMA is applied to the intermediate wafer, the residual resist will be fully baked out post annealing.[6] We thus only apply a brief $O_2$ descum to clean the surface. However, if AZ 1505 is used instead, the photoresist will crosslink post baking, which requires boiling di-acid cleaning (1:1 $H_2SO_4$:$HNO_3$ for 2 h at the nitric boiling point). Due to the much reduced viscosity of AZ 1505 compared with PMMA, we applied PMMA-based bonding on structured surfaces and acid-sensitive substrates, while using AZ 1505 on other substrates. The final device is shown in Figure S2 (e). We note that our bonded heterostructure is compatible with isopropyl alcohol, acetone, potassium or tetramethylammonium hydroxide (TMAH) based developers (such as AZ 300 MIF or AZ 400K), heated (80 °C) N-Methyl-2-pyrrolidone (NMP), and room temperature NanoStrip. However, the tri-acid cleaning (1:1:1 $H_2SO_4$:$HNO_3$:$HClO_4$ at refluxing temperature), hot (≥80 °C) Piranha (3:1 $H_2SO_4$:$H_2O_2$), and hot NanoStrip may damage the bonds and loosen the membranes from the target wafer.

# 2 Material characterizations

## 2.1 Out-of-plane strain of smart-cut membranes

Unlike isotropically-etched diamond frames[7] or ICP-etched diamond slabs[8], smart-cut diamond membranes naturally contain out-of-plane strain originated from the lattice mismatch between the damaged layer generated from He$^+$ implantation and the subsequent overgrowth layer. This strain brings a curvature to freestanding membranes due to their high geometry aspect ratio (usually beyond 500), and has been observed in previous works[9]. To roughly estimate the strain magnitude, we performed Raman spectroscopy on a transferred diamond membrane with 100 nm overgrowth layer prior to ICP etching. The experimental data is shown in Figure S3 (a) as individual points, which can be fit by two Lorentzian curves. The original membrane (He$^+$ damaged layer) is indicated as the dashed blue curve with a center wave number of 1326 cm$^{-1}$, while the overgrowth layer obtains a center wave number of 1332 cm$^{-1}$, labelled as the dashed orange curve. Differences in wave number indicate a ≈0.5 % lattice mismatch.

In Figure S3 (b) a test membrane partially attached to a PDMS2-stamp is shown, with the upper and lower parts floated, as pointed by the red arrow. From the interference pattern we can observe the extension of the original layer and the compression the overgrowth layer, leading the membrane to be curved up. The strain elimination via ICP etching is discussed in section 1.5.

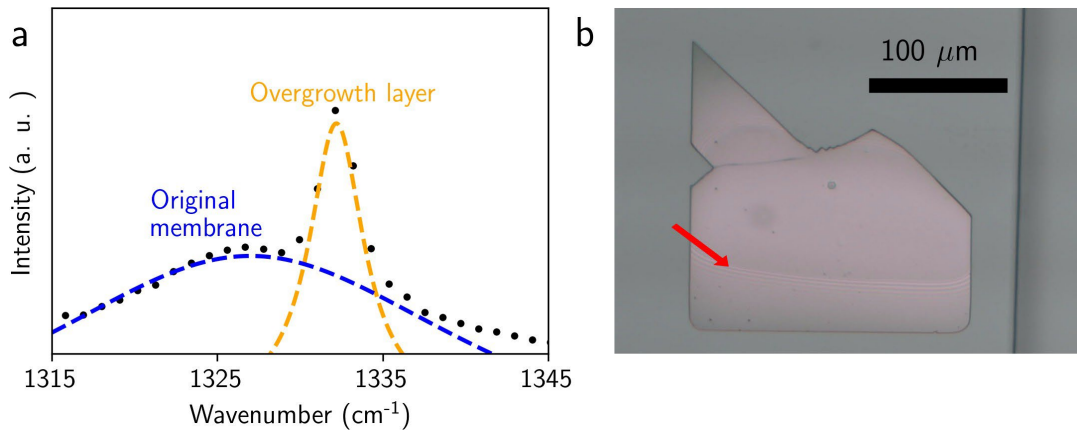

Figure S3: The out-of-plane strain in smart-cut diamond membranes. (a) Raman spectroscopy of a transferred membrane prior to the ICP etching. The raw data (black dots) can be qualitatively fitted by two separate peaks, the damaged (dashed blue line) and the overgrown (dashed orange line) layers. (b) A microscope image of a curved membrane on a PDMS2-stamp. The interference pattern pointed by the red arrow originates from the airgap between a curved membrane and a flat PDMS surface.

## 2.2 Surface morphology of diamond and final wafers

The success of plasma-enhanced bonding highly depends on the surface morphology of the diamond membrane and target substrates. We performed AFM to characterize the surface roughness along the fabrication process. Both small (200 nm by 200 nm) and large (10 μm by 10 μm) scale scans were applied to capture features of various sizes.

For diamond membrane surfaces, we only discuss the etched side in this section since the growth side preserves the excellent surface morphology as discussed in our previous work[1]. Prior to the ICP etching, the etched side has an $R_q$ of 1.17 nm due to the He$^+$ implantation and EC etching. By applying 15 cycles of Ar/Cl$_2$ etching discussed in section 1.4, we show a polishing effect with $R_q$ reduced to 0.54 nm (0.44 nm) in small (large) area scans, as shown in Figure S4 (a)-(b). We note that small area scans usually reveal greater roughness compared to large area scans, which might indicate the Cl-based contamination on the diamond surface as discussed in previous studies.[8] Such contamination can be removed by O$_2$/Cl$_2$-O$_2$ ICP cycles, shown as an $R_q$ of 0.25 nm (0.34 nm) in small (large) areas (Figure S4 (c)-(d)). We did not observe the change of $R_q$ post O$_2$ descum treatment (0.28 nm and 0.34 nm in small and large area scans), as depicted in Figure S4 (e)-(f). In contrast, our customized plasma recipe is found to have a negative impact on the surface morphology by elevating the $R_q$ to 0.84 nm (1.09 nm) in small (large) areas. This can be interpreted as an appearance of particle-like dust since the $R_q$ of the contamination-free area remains to be ≤0.35 nm. Such contamination can be reduced by transitioning to process specific tooling.

We also analyzed the impact of downstream plasma asher recipes on two widely used carrier substrates, namely fused silica and thermal oxide silicon wafer. Values of $R_q$ are shown in Table S1. Both wafers exhibit sub-nm $R_q$ out of the box, and their surface morphology is maintained post ashing with no correlation to power or duration settings. Therefore, we conclude that our plasma recipe has no significant effect on the surface morphology of these carrier wafers.

| Carrier wafer type | AFM area | No plasma | O$_2$ descum | Custom O$_2$ plasma |
|---|---|---|---|---|
| Fused silica | 200 nm x 200 nm | 0.324 nm | 0.281 nm | 0.366 nm |
| | 10 μm x 10 μm | 0.685 nm | 0.581 nm | 0.528 nm |
| Thermal oxide | 200 nm x 200 nm | 0.257 nm | 0.290 nm | 0.352 nm |
| | 10 μm x 10 μm | 0.270 nm | 0.427 nm | 0.293 nm |

Table S1: $R_q$ of fused silica and thermal oxide wafers under various plasma recipes.

## 2.3 Height variation across diamond membranes

### 2.3.1 One dimensional (1D) height detection via profilometry

In this work two methods are applied for global flatness characterizations. For 1D characterization presented in the main text, we use a profilometer (Dektak XT) to scan across the membrane. The scan range is 350 μm with a 6.5 μm height detection limit. The membrane shows a height variation $\sigma$ of ≈1 nm, which is 10 times smaller than our HSQ-bonded membranes' value mea-sured on the same equipment[1]. This $\sigma$ is also below the minimum detectable height of the tool (10 nm) and the instrument resolution (1.5 nm) for large scale scanning.

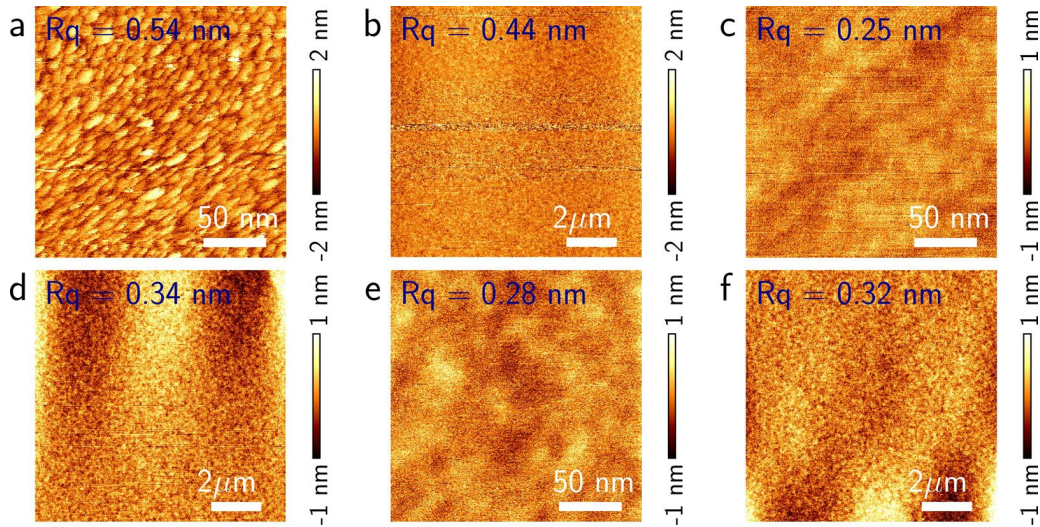

Figure S4: Small (200 nm scanning range) and large (10 µm scanning range) area AFM of the etched side of diamond membranes under various plasma conditions. (a)-(b) Post 15 Ar/$Cl_2$ cycles with recipe described in section 1.4. (c)-(d) Post additional 3 $O_2$/$Cl_2$-$O_2$ cycles with recipe described in section 1.4. (e)-(f) Post $O_2$ descum recipe described in section 1.5. The global patterns in (d) and (f) reflects the flattening fitting of the resist layer with height variation which is not related to the membrane surface morphology.

### 2.3.2 Two dimensional (2D) height mapping via confocal laser scanning microscopy (CLSM)

The 2D membrane height map and surface topology is measured via an Olympus LEXT OLS4100 405 nm laser confocal microscope. The microscope image of the measured membrane-thermal oxide silicon heterostructure is shown in Figure S5 (a) with its height map shown in Figure S5 (b). The bonded membrane profile reveals a uniform height of 309±8 nm across the membrane, with the standard deviation $\sigma$ below the height resolution of the CLSM ($\approx$10 nm). Currently, the dominant sources of height inhomogeneity are assigned to diamond membrane crystallographic growth defects[10] and transfer process contamination, which can be minimized by performing the totality of the processing in a clean environment (e.g. cleanroom).

## 2.4 Hydrophilicity characterization of bonding interfaces

### 2.4.1 Contact angle measurement setup

We measure water contact angle to characterize the surface hydrophilicity of diamond and target substrates. Measurements were performed using a Kruss DSA100A dropped shape analyzer. DI water was dispensed from a sterile syringe (14-817-25, Fisher Scientific) through a thin needle (75165A761, McMaster-Carr). The dispense rate was set to 2.67 µL s$^{-1}$, resulting in a typical droplet size between 4 µL to 5 µL. The diamond we used for contact angle measurements are 3 mm by 3 mm single crystal fine-polished diamond substrates ($R_q \leq 0.3$ nm). Figure S6 (a)-(b) show the contact angle analysis of diamond and thermal oxide substrates before and after high power plasma ashing, indicating an improvement of the surface hydrophilicity. We note that the weaker $O_2$ descum recipe showed minimal influence on diamond hydrophilicity, with the contact angle above 40° (not shown in the figure).

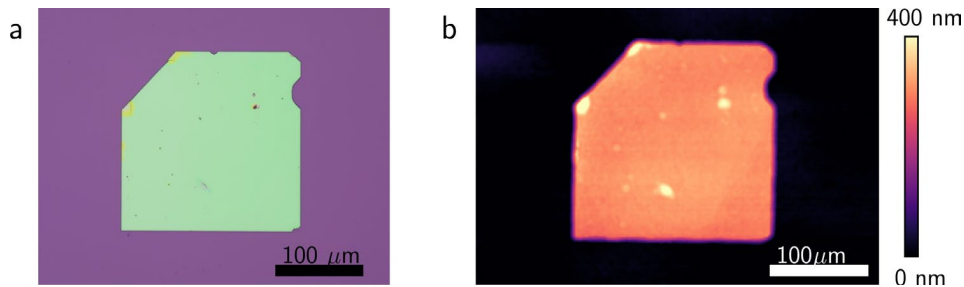

Figure S5: CLSM of a membrane bonded to a thermal oxide silicon substrate. (a) The microscope image of the characterized heterostructure. Several growth defects are present on this specific diamond membrane. (b) The 2D height map of the diamond membrane, showing an average height of $309 \pm 8$ nm. This data had a plane-fit adjustment performed to remove any substrate tilt aberrations. The X-Y resolution is quoted as $\leq 0.2$ µm, whereas the Z resolution is hardware defined to $\leq 10$ nm.

### 2.4.2 Aging and temperature dependence of the surface hydrophilicity

As stated in previous studies,[11] surface hydrophilicity is positively correlated to the plasma-activated bonding quality. Here we characterize the decay of the hydrophilicity by recurring measurements of contact angle on various substrates, including diamond, fused silica, thermal oxide, sapphire and lithium niobate on insulator. The aging trend of the hydrophilicity is shown in Figure S6 (c), indicating the need for a timely bonding process. We also tested the temperature dependence of the hydrophilicity by baking the plasma-treated diamond sample on a 90 °C hotplate for 30 s prior to the contact angle measurements. A decay of hydrophilicity was observed, as shown in Figure S6 (d), possibly due to the loss of surface-absorbed water molecules. Partially due to the strong association between elevated temperature and reduced hydrophilicity, we chose resist AZ1505 as one of the mounting media in our bonding process for its much reduced viscosity at a fairly low glass transition temperature (softening temperature), see section 1.3.

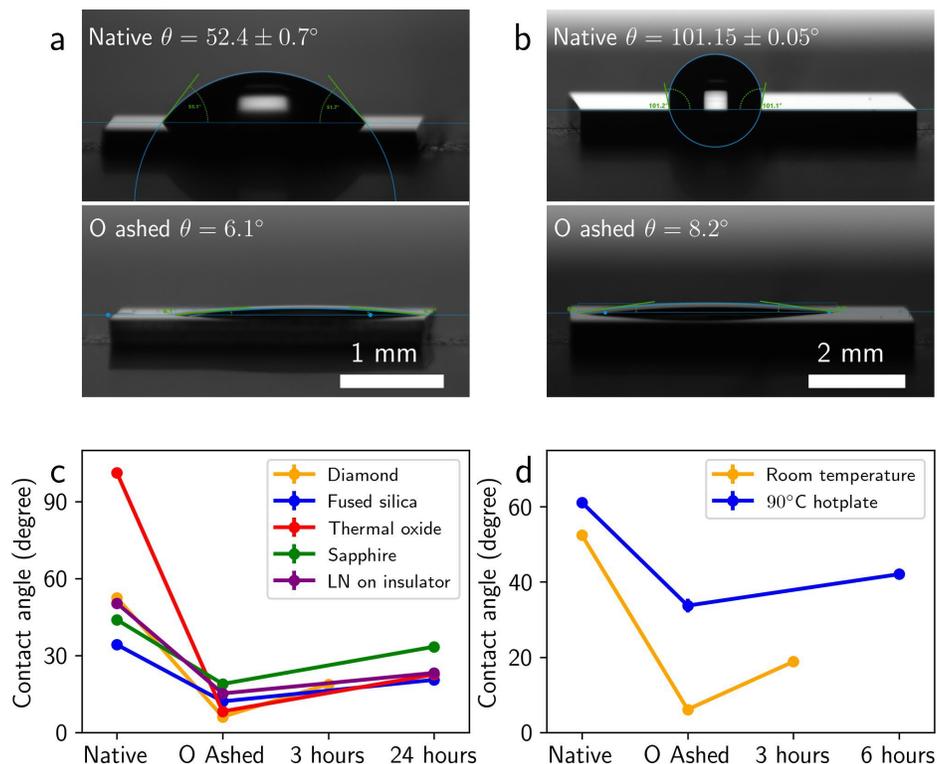

Figure S6: Surface hydrophilicity characterization via contact angle measurements. (a)-(b) Contact angles of native (top) and high power plasma-treated (bottom) diamond and thermal oxide substrates. (c) Aging of hydrophilicity on various substrates. (d) Hydrophilicity of diamond surface with (blue) or without (orange) 30 s baking on a 90 °C hotplate.

## 2.5 High resolution transmission electron microscopy (HRTEM) and Energy-dispersive X-ray spectroscopy (EDS)

To obtain an atomic level understanding of the bonding interface, we performed HRTEM on a cross-sectional sample from a diamond-sapphire heterostructure. The sapphire substrate is C-axis (0001) from University Wafer. To start, a 200 nm-thick gold mask was deposited on the surface to protect the diamond membrane being damaged by Ga ion beam. Using a Zeiss NVision 40 system, a cross-sectional TEM specimen with thickness of a few tens of nanometer was prepared by standard FIB lift-out procedure. The HRTEM image was obtained by a FEI Titan operated at 200 kV, which was equipped with aberration corrector and chromatic corrector. The scanning transmission electron microscope (STEM) image was acquired by using high-angle annular dark field (HAADF) detector, as shown in Figure S7 (a). A FEI Talos S/TEM equipped with a Super X energy-dispersive spectrometer (EDS) was employed for STEM-EDS elemental mapping. The result is shown in S7 (b)-(d).

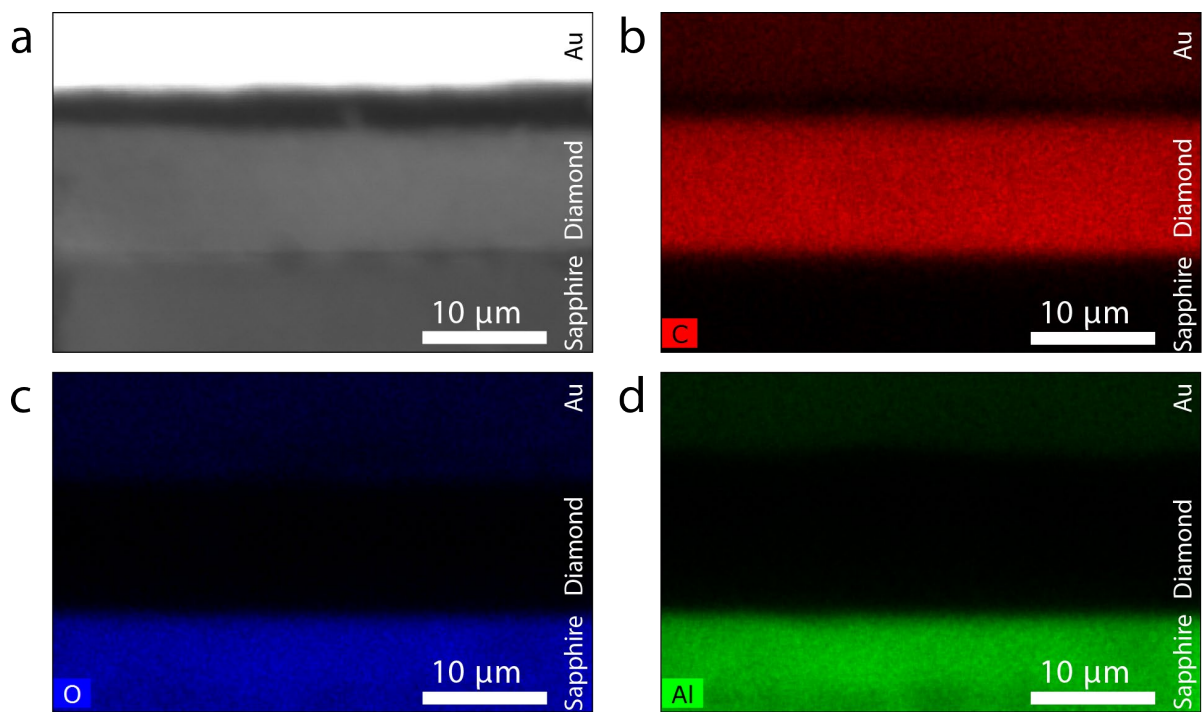

Figure S7: Additional atomic scale analysis of the bonded membrane. (a) HAADF-STEM of the diamond-sapphire heterostructure. (b)-(d) STEM-EDS elemental analysis of the diamond-sapphire heterostructure. The intensity of carbon, oxygen and aluminum elements at the cross section is shown in (a), (b) and (c), respectively.

## 2.6 X-ray photoelectron spectroscopy (XPS)

The experimental samples were identical to those employed in the rest of the demonstrations: diamond, fused silica, and sapphire. A set of these was left unprocessed to be used as a reference, whereas the other set received surface activation via oxygen plasma ashing ≈90 min before loading into the XPS chamber. Two different incidence angles were taken for the XPS analysis to confirm that the perceived effect was related to near-surface species. One set of characterization was taken at 0° incidence, with another set taken at 35° incidence. No significant differences were found for both datasets, as such only the 0° incidence dataset is shown. In agreement with the contact angle measurements, in that timeframe, the surface is known to have degraded somewhat, however it remains 'bond-ready'. As such, all the quantitative XPS analysis provides a lower bound on the surface activation related species. Future experiments will be necessary to study the 'as-ashed' surfaces.

An Al-K$\alpha$ source in a Thermo Scientific ESCALAB 250Xi was used to perform the XPS characterization at both normal and 35° incidence angles. Elemental peaks were taken with a pass energy of 50 eV, 50 ms dwell time, and a step size of 0.1 eV, whereas the C KLL peak had 100 eV pass energy and 0.5 eV step size. For all scans, a charge compensating electron flood gun was used whereas the X-ray spot size was maintained at ≈200 µm in spread. C 1s, C KLL, Al 2p, Si 2p, O 1s, N 1s peaks were all collected in high-resolution mode and the presence of other unplanned contaminants was verified via survey scans. All scans were references to the C sp$^3$ peak at 284.8 eV binding energy (BE). Elemental analysis was taken by quantitative comparison of the high-resolution components fits (with appropriate sensitivity factors accounted for). The C 1s KLL signature was used to extrapolate and corroborate sp$^2$-to-sp$^3$ ratio obtained from the C 1s line-fit by calculating and fitting the peak-to-peak separation of the first derivative of the KLL signal[12,13]. All peak fitting parameters were taken from commonly agreed on values for equivalent materials and systems from the NIST XPS database.[14] XPS high-resolution spectra of relevant lines with fitting parameters are shown in Figure S8 and Figures S17, S18, S19, S20, and S21 in the section 6 for the Unprocessed-Diamond, Ashed-Diamond, Unprocessed-Sapphire, Ashed-Sapphire, Unprocessed-Silica, and Ashed-Silica, respectively. The analysis is quite consistent with the conclusions of this paper. However, an unresolved abnormality needs to be pointed out, specifically the presence of an unidentified third component of the O 1s peak in the fused silica samples. While currently that peak is identified as organics, this is a placeholder title. Looking at the appearance of an intermediate $SiO_2$-Silicate peak in the fused silica Si 2p quantification, it is plausible that these are the oxygen species from low coordination quartz. However, at this time further experiments would need to confirm this hypothesis. Regardless, these are in the low at.% range ($\approx 2$ at.%) and do not impact any of the conclusions of this paper. Table S2 presents the full quantification summary of the XPS experiments.

| Peak - Component | XPS Quantification (at.%) | | | | | |
|---|---|---|---|---|---|---|
|  | Diamond Raw | Diamond Ashed | Sapphire Raw | Sapphire Ashed | Silica Raw | Silica Ashed |
| C 1s - sp3 | 75.6 | 77.9 | | | | |
| C 1s - sp2 | 6.5 | 5.9 | 5.9 | 5.4 | 3.7 | 2.0 |
| C 1s - C-O-C | 8.6 | 8.2 | 0.3 | 0.5 | 1.7 | 0.8 |
| C 1s - C=O | 1.6 | 1.3 | 1.0 | 0.8 | 0.8 | 0.5 |
| C KLL - sp2 | 12.8 ± 3 | 4.5 ± 3 | | | | |
| O 1s - C=O | 3.3 | 2.0 | 4.5 | 8.4 | | |
| O 1s - C-O | 4.2 | 4.5 | | | | |
| O 1s - Al/Si Oxide | | | 40.0 | 38.3 | 51.9 | 53.8 |
| O 1s - Substochio. | | | 3.4 | 3.9 | 0.9 | 0.7 |
| O 1s - Un-ID. | | | | | 2.1 | 2.6 |
| Al 2p - Oxide | | | 42.0 | 39.7 | | |
| Al 2p - Substochio. | | | 2.8 | 3.0 | | |
| Si 2p - SiO2 | | | | | 16.6 | 18.7 |
| Si 2p - Un-ID. | | | | | | 1.6 |
| Si 2p - Silicate | | | | | 22.1 | 18.9 |
| N 1s - Adsorbed N | 0.2 | 0.1 | 0.2 | 0.1 | 0.3 | 0.2 |

Table S2: High-resolution XPS quantification for all the fitted components and peaks.

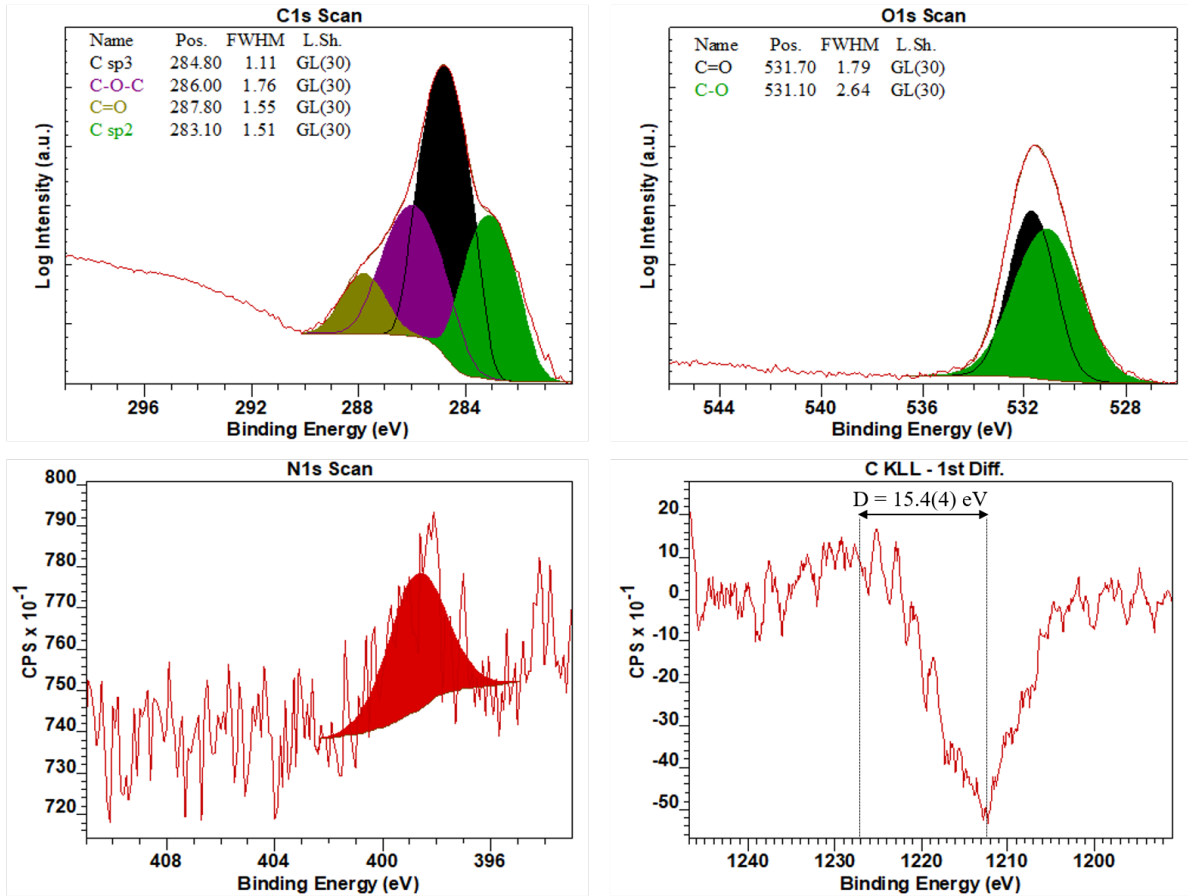

Figure S8: XPS characterization of the unprocessed diamond substrate, showing the C 1s and O 1s deconvolutions with component labels and fitting parameters. sp$^2$ content quantification is achieved via D parameter linear extrapolation of the C KLL first order derivative[12,13]. N 1s peak contamination also showed as samples were exposed to atmosphere. Some of the peak fittings are presented in log scale to help discern minority components.

# 3 Optical characterizations of GeV$^-$ and NV centers

## 3.1 Optical setup

In this work, membrane samples are mounted inside a closed-loop cryostation (Montana S200) and cooled down till 4 K for low temperature measurements. The position of the membrane is controlled by three closed-loop piezo micropositioners (Attocube ANC 350). Light beams are navigated by a fast steering mirror (Newport FSM-300). For photoluminescence (PL) measurements, we either use a 519 nm green diode (Thorlabs LP520-SF15) or a 532 nm continuous wave (CW) laser (Lighthouse Photonics Sprout-G) as the excitation source. For photoluminescence excitation (PLE) measurements of GeV$^-$, the excitation laser is generated by a wave mixing module (AdvR Inc.) combining a tunable CW Ti:Sapphire laser (M Squared Solstis) and a monochromatic CW laser (Thorlabs, SFL 1550P). A single photon counting module (SPCM) (Excelitas Technologies) is applied to plot PL maps, while a spectrometer (Princeton Instruments, SpectraPro HRS) is used to measure the spectra of the color centers. We combine two bandpass filters (Semrock FF01-615/24-25, Semrock FF01-600/14-25) for GeV$^-$ PL measurements, and use bandpass filters (2× Semrock FF01-647/57-25) for PLE measurements. For NV measurements, a single long pass filter (Semrock LP02-561RE-25) is used to capture both NV$^0$ and NV$^-$ signals.

## 3.2 Effect of plasma treatments on the optical coherence of GeV$^-$ centers

As mentioned in section 1.5 and 2.2, oxygen plasma on diamond membranes is optional for the bonding process. In this section, we investigate the effect of the plasma on the PL and PLE properties of GeV$^-$ centers. Two membranes, each has ≈200 nm thickness and contains 40 nm-deep implanted GeV$^-$ centers from the top surface, were transferred on a single thermal oxide wafer. The membrane 1 received no plasma treatment prior to the bonding, while the membrane 2 received a strong plasma ashing introduced in the section 1.5. Figure S9 (a)-(b) shows the 4 K PL map of the GeV$^-$ centers on membrane 1 and 2, which clearly indicates a signal-to-background improvement via plasma treatments. The average background dropped from ≈7000 to ≈1900, leading to an improvement of signal-to-background ratio from ≈4.5 to ≈11. The slightly lower signal shown in Figure S9 (b) indicates a slight oxygen termination which shifts the Fermi level away from the optimal value for GeV$^-$ centers. Figure S9 (c) shows the single (2.5 min average) ZPL linewidths with resonant excitation. We observed no statistical difference of the linewidth distribution, with mean single scan linewidth of 97 MHz (85 MHz) and mean average scan linewidth of 212 MHz (196 MHz) for membrane 1 (2). The measured linewidths are broader than the real value due to the resolution limit of the wavelength meter (High Finesse WS6-600, 20 MHz measurement resolution, 500 MHz wavelength accuracy). A separate characterization of tin vacancy centers in our membranes using a higher resolution wavelength meter reported a much narrower linewidth profile, with the average linewidth only ≈50 % higher than the transform-limited value[15].

## 3.3 Strain characterization via GeV$^-$ centers

Group IV centers in diamond are good sensors for local strain environment due to their relatively large strain susceptibilities[16]. The strain magnitude can be estimated via the relative shift of the wavelength $|\epsilon_{A1g} - \epsilon_{A1u}|$ and the increased ground state splitting $2\sqrt{\epsilon_{Egx}^2 + \epsilon_{Egy}^2 + \left(\frac{\lambda_{SO}}{2}\right)^2}$.

We recorded spectra from 69 (52) GeV– centers in diamond membranes direct-bonded to fused silica (thermal oxide) carrier wafers. The ZPL wavelength and ground state splitting distribution are shown in Figure S9 (d). Since implantation-induced strain has a large span which greatly affects GeV$^-$'s ZPL and ground state splitting statistics, we only focus on the color centers with ZPL from 601.5 nm to 603.2 nm and ground state splitting ≤800 GHz to estimate the strain from bonded membrane crystals. This region covers ≈80 % of our data points.

For membranes bonded to fused silica (thermal oxide) wafers, the average ZPL wavelength of GeV$^-$ centers is 602.68(20) nm (602.53(8) nm), with the average ground state splitting to be 307(158) GHz (224(75) GHz). These ZPL wavelength distributions are comparable with those obtained in bulk diamonds.[17] We do observe a slight positive strain with diamond-fused silica heterostructures, which could be explained by the lower thermal expansion ratio of fused silica. Thermally induced negative strain is barely visible for diamond membrane-thermal oxide substrates, which may come from the fact that membranes with such a high aspect ratio ($\geq 1000$) could deform instead of generating negative strain under compressive stress. We estimate the average strain level of diamond membranes to be $\approx 2.9 \times 10^{-4}$ ($\approx -1.7 \times 10^{-4}$) on fused silica (thermal oxide) carrier wafers.

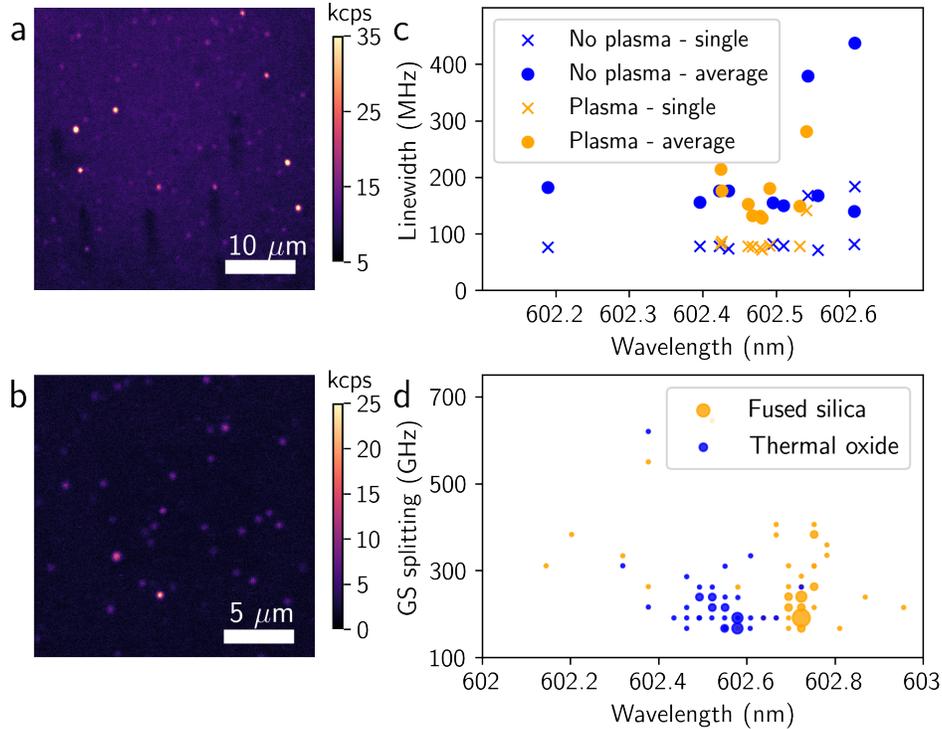

Figure S9: Additional optical characterization of GeV⁻ centers in bonded membrane at 4 K. (a) A PL map of GeV⁻ centers with no plasma treatments prior to the bonding. (b) A PL map of GeV⁻ centers with diamond treated via high power plasma recipe. (c) ZPL wavelength and single/2.5 min average scan linewidth distribution of GeV⁻ centers. Centers from native (plasma treated) membranes are labelled in blue (orange). (d) ZPL wavelength and ground state splitting statistics of GeV⁻ centers in diamond membranes bonded to fused silica (orange) or thermal oxide (blue) substrates. The size of each disk reflects the occurrence of the same data point within the resolution of the spectrometer ($\approx 0.04$ nm or $\approx 30$ GHz). This resolution also accounts for the artificial, equally spaced data pattern shown here.

### 3.4 NV centers at 4 K

Thanks to the low optical background of the direct bonding method, we were able to resolve individual NV centers in diamond membrane heterostructures. A typical NV PL map taken at 4 K is shown in Figure S10 (a), with a signal-to-background ratio of over 1.4. This paves the way of towards NV sensing applications, as discussed in the main text. In addition, the charge stability of NV centers is a good indicator of membrane's surface termination with respect to various plasma treatments on the diamond bonding interface. Here we characterized the NV spectra in three bonded membranes. They were picked up from a single mother substrate doped *in-situ* with $^{15}$N, thus contain same NV densities. They were bonded to $SiO_2$ surfaces with (1) no plasma treatment (2) $O_2$ descum (3) high power plasma ashing on the diamond bonding side. The typical NV spectra and statistical data is shown in Figure S10. We observe that with no plasma treatment, the NV center stays at the neutral charge state due to the hydrogen termination effect of the $Ar/H_2$ annealing process.[18] However, membranes treated with plasma have considerably higher $NV^-/NV^0$ ratio. This ratio is positively correlated with the strength of the $O_2$ ashing process, indicating a better oxygen termination performance which helps maintain the NV center in its negatively charged state[19]. Indeed, it is expected that such a dry O-termination method can be used to desirably engineer the diamond near-surface Fermi level for $NV^-$ based applications[20,21]. Systematic characterizations of the effect of such plasma activation processing on near-surface $NV^-$ are ongoing.

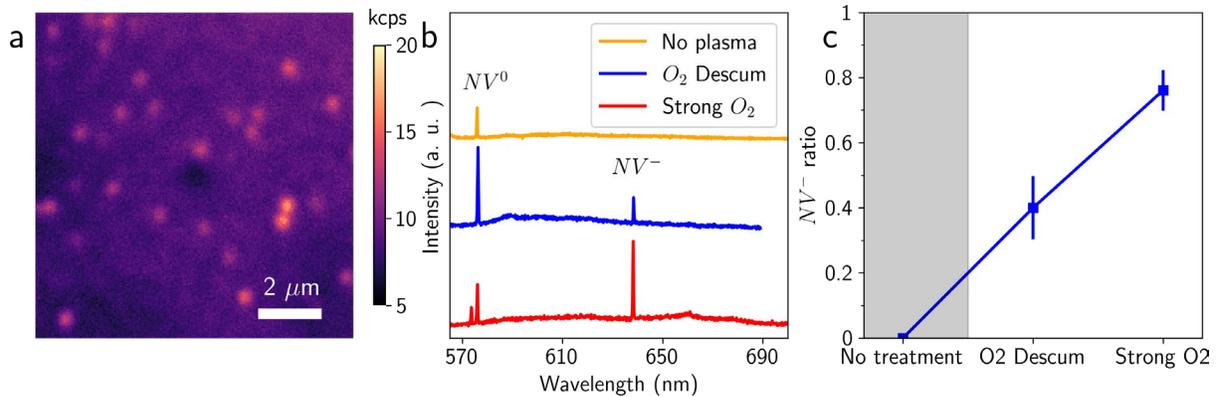

Figure S10: Additional optical characterization of NV centers in direct-bonded diamond membrane measured at 4 K. (a) A PL map of NV centers in a diamond membrane with 325 μW green excitation and 561 nm longpass filter. The signal-to-background ratio is ≈1.4. (b) The spectra of NV centers with no plasma (orange), $O_2$ descum (blue) or high power $O_2$ plasma treatment (red) prior to bonding. The dominant $NV^0$ originates from the $Ar/H_2$ annealing. The appearance of $NV^-$ indicates the surface oxygen termination of the diamond bonding interface. The small peak on the left side of the red curve is the Raman response of diamond due to the different excitation laser wavelength (532 nm instead of 519 nm). (c) The ratio of NV in its negatively charged state with respect to different plasma conditions on the diamond bonding interface. The $NV^-$ ratio increased from 0 to 76(6) %. Calculation of this ratio takes the Huang-Rhys factor of $NV^0$ (3.3) and $NV^-$ (4.0) into account.[22]

# 4 Nanophotonic cavities with direct-bonded diamond membranes

## 4.1 Optical setup for nanophotonic cavity measurements

A similar confocal setup as the 4 K one is used for cavity measurements, but the system operates at room temperature. Two separate beams controlled by galvanometer systems (Thorlabs GVS102) are applied to perform transmission spectroscopy. A pulsed supercontinuum source (430 nm to 2400 nm, SC-OEM YSL Photonics) is used for broadband excitation, and a spectrometer (Princeton Instruments, SpectraPro HRS) is used to observe the optical bandgap and measure the cavity resonance. Finer cavity linewidth measurements are performed by scanning a Ti:Sapphire laser (M Squared Solstis) around resonant wavelengths of the cavities. Transmission signal is collected by a SPCM (Excelitas Technologies).

## 4.2 TiO$_2$-based nanophotonic structures

### 4.2.1 Fabrication process of TiO$_2$ cavities

Our TiO$_2$ nanofabrication process is similar with the one described in previously.[23] The process starts with a 340 nm electron beam resist deposition (950 K PMMA A4, MicroChem), followed by a 20 nm gold deposition via thermal evaporation (Nexdep PVD platform, Angstrom Engineering) to prevent charging effects. Cavity patterns are lithographically defined at a dose of 1200 μC cm$^{-2}$ at 100 keV (Raith EBPG5000 plus), and the gold layer is removed by TFA gold etchant (Transene) afterwards. Exposed patterns are developed in a 1:3 MIBK:IPA solution on a cold plate at 7 °C for 90 s, followed by a 60 s of IPA stopper and 60 s of DI water rinse. A 3 s ICP RIE etching is then conducted to remove the residual resist post developing. TiO$_2$ deposition takes place in an atomic layer deposition (ALD) system (Savannah Thermal ALD, Veeco) at 90 °C. The precursors of the TiO$_2$ deposition include tetrakis (dimethylamido) titanium (TDMAT) and water. The deposition thickness is 200 nm to 300 nm, depending on the actual geometry of the structure. Overfilled TiO$_2$ is removed by another ICP RIE etch and the remaining resist is stripped by nanostrip (MicroChem). Finally, the structures are annealed at 250 °C on a hotplate for 2 h to improve their optical quality. Scanning electron microscopy (SEM) images of fishbone cavities and ring resonators on the diamond membrane are shown in Figure S11 (a) and (b), featuring planarized top surfaces and smooth sidewalls.

### 4.2.2 Optical characterization of TiO$_2$ ring resonators

We measured the transmission of the TiO$_2$-diamond ring resonator through the drop port. The resonant excitation data for transverse electric (TE) and transverse magnetic (TM) modes are identified via separate polarization measurements. We observed high quality factors $Q$ for both modes, with $Q_{TE}$ = 12620 and $Q_{TM}$ = 16320. The spectrum of the ring resonator and the resonant scan results are shown in Figure S11 (e). We note that this is the first demonstration of deposited TiO$_2$ ring resonators on diamond. Our current ring resonator design supports both modes, but the quality factor for either mode can be further improved via geometry optimization.[24]

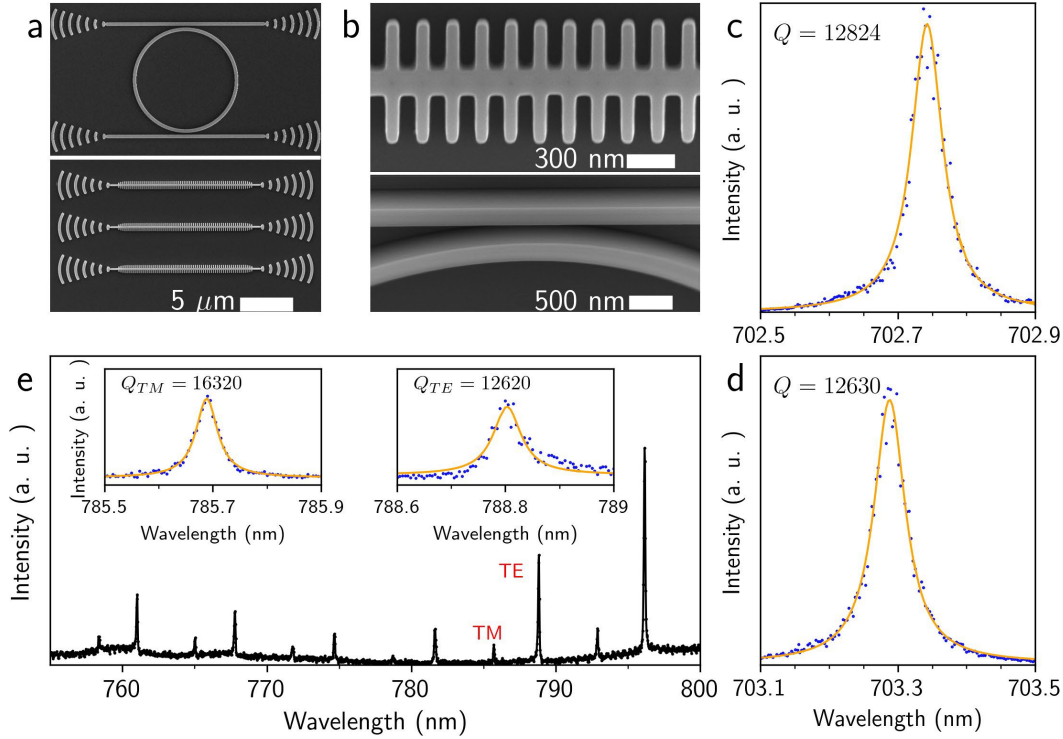

Figure S11: Additional information of TiO$_2$ nanophotonic devices. (a) Zoomed-out SEM images of fishbone cavities and ring resonators on diamond membrane. (b) Zoomed-in SEM images of fishbone cavities and ring resonators, featuring flat top surfaces and smooth sidewalls. (c-d) Quality factors $Q_{TiO_2}$ of two best TiO$_2$ fishbone cavities directly fabricated on bare fused silica substrates. (e) The transmission spectrum of the TiO$_2$-based ring resonator measured at the drop port. Insets: the TE and TM cavity resonances measured with a tunable laser as the excitation source.

### 4.2.3  TiO$_2$ cavities on fused silica

To estimate the optical loss of the direct bonded diamond heterostructure, we repeated the TiO$_2$ fishbone cavity fabrication on bare fused silica substrates and measured their optical transmission via resonant excitation. The maximum quality factor $Q_{TiO_2}$ we measured is 12 824, with the average $Q_{TiO_2}$ of the two best cavities is 12 727, as shown in Figure S11 (c) and (d). This $Q_{TiO_2}$ includes the optical scattering within the deposited TiO$_2$ and the surfaces of the cavity (top surface, interface between TiO$_2$ and fused silica, sidewalls etc.). Subtracting the average $Q$ of TiO$_2$ cavities on diamond membrane returns an approximate optical loss of the bonded diamond-fused silica system, which is $Q_{sys} \approx 50000$. This system optical loss includes the optical scattering from the diamond/fused silica interface, the diamond/TiO$_2$ interface and the diamond crystal itself, instead of the loss from the TiO$_2$/fused silica interface.

## 4.3 Fabrication process of diamond cavities with bonded membranes

Diamond membranes are thinned down to 280 nm and then bonded to a thermal oxide substrate with 1 µm oxide thickness for diamond-based nanophotonic structures. We first deposit a hard mask of 25 nm alumina via ALD (Savannah Thermal ALD, Veeco), then spin a 90 nm of electron beam resist (ARP6200.04, MicroChem), and thermally evaporates 20 nm of gold (Nexdep PVD platform, Angstrom Engineering) to prevent charging effects. Cavity patterns are lithographically defined at a dose of 250 µC cm$^{-2}$ at 100 keV (Raith EBPG5000 plus), and the gold layer is removed by TFA gold etchant (Transene) afterwards. Exposed patterns are developed in a Amyl Acetate for 60 s, followed by a 60 s of IPA stopper and 60 s of DI water rinse. A 30 s alumina ICP RIE etching is then conducted to transfer the pattern into the hard mask. This is followed by resist removal in heated NMP at 80 °C. We then conduct multiple cycles of ICP RIE etching with O$_2$/Cl$_2$-O$_2$ plasma in 15 s intervals till the diamond is etched all the way through. Lastly, we performed a 20 s of Al$_2$O$_3$ etching to remove the hard mask.

# 5 Heterostructure-enabled flow channel for molecular sensing applications

## 5.1 Widefield microscope setup

A custom-built widefield fluorescence microscope in an inverted configuration operated at room temperature was used to acquire all the widefield images presented in this section. It is equipped with a 60× oil objective (Olympus UPLAPO60XOHR) with $NA$ = 1.5. A motorized translational stage allows the displacement of the focus point of the laser beam on the back focal plane of the objective, enabling a continuous change in excitation beam angle for episcopic and total internal reflection fluorescence (TIRF) illumination. Two independent laser sources (488 nm and 532 nm, 100 mW maximum power output, Coherent Sapphire) are available for excitation, and an Andor iXon Ultra 888 electron-multiplying charge-coupled device (EMCCD) is used for image collection. Unless otherwise stated, EMCCD was cooled down to −60 °C to minimize noise. We implement an iris in the excitation path to manually adjust the size of the laser beam. A Hamamatsu W-VIEW GEMINI system is placed in front of the EMCCD, which is usually operated in the bypass mode unless taking dual-imaging data.

## 5.2 Imaging of NV$^-$-hosting diamond membranes

### 5.2.1 Widefield imaging

We bonded diamond membranes containing $\delta$-doped NV$^-$ to the center of 25 mm by 8 mm rectangular-shaped fused silica slabs diced from 2 inch wafers (Part#: U01-131121-1 from Uni-versity wafer, 180 µm thick). One of these samples was characterized thoroughly as discussed thereafter. For simplicity, we refer this particular sample as Sample A. Diamond membranes from the same mother substrate as Sample A have been characterized previously[1]. We applied immersion oil (Olympus Type F) to the backside of the coverslip (opposite to the bonded diamond membrane) and imaged through it. A long-pass optical filter (Semrock LP594) was used. The maximum excitation power of the 532 nm laser when the beam size matches the effective field-of-view (73 µm by 73 µm) was ≈35 mW, namely a power density of approximately 800 W cm$^{-2}$. We note that a typical confocal setup for NV$^-$ experiments operates with a power density of approximately 10 kW cm$^{-2}$ to 100 kW cm$^{-2}$, which is 1 to 2 orders of magnitude higher than our widefield imaging system. The micrograph shown in S12 (a) is a rotated version of Figure 4 (a) in the main text, which was acquired with 100 s exposure time and 300 gain. We use an iris to restrict the size of excitation beam to ≈35 µm in diameter, which avoids the direct illumination of the membrane edges. The fluorescence background of the diamond membrane region was very low, which is comparable and even appeared to be darker than the fused silica coverslip due to the large refractive index of diamond. The bright, diffraction-limit spots in the diamond membrane, later verified as predominately NV$^-$ centers, showed excellent photostability under continuous excitation (would not be photobleached).

### 5.2.2 Confocal imaging and NV$^-$ verification

We performed a confocal scan on Sample A shown in Figure S12 (a) and confirmed that the observed bright emitters in widefield image were mostly NV$^-$ centers. This verification was done by measuring the optically detected magnetic resonance (ODMR) feature of NV$^-$ centers with external microwave signals. The confocal scan and the continuous wave ODMR (CW-ODMR) were performed on a custom-built confocal microscope equipped with a fast steering mirror and a 520 nm laser as excitation (70 µW). Microwave signal was generated from a vector signal generator (Stanford research systems, SG394) and delivered to the sample through a 25 µm-thick gold wire. Pulse control was achieved using a pulse streamer (Swabian Instruments Pulse Streamer 8/2). All measurements were done at room temperature with no external magnetic field. We characterized 14 bright emitters in a 10 µm by 5 µm area as shown in Figure S12 (f). The ODMR scanning range was between 2.82 GHz to 2.92 GHz. As the result, 12 out of 14 emitters showed the characteristic NV$^-$ ODMR feature centered at 2.87 GHz with 2 % to 6 % signal contrast.

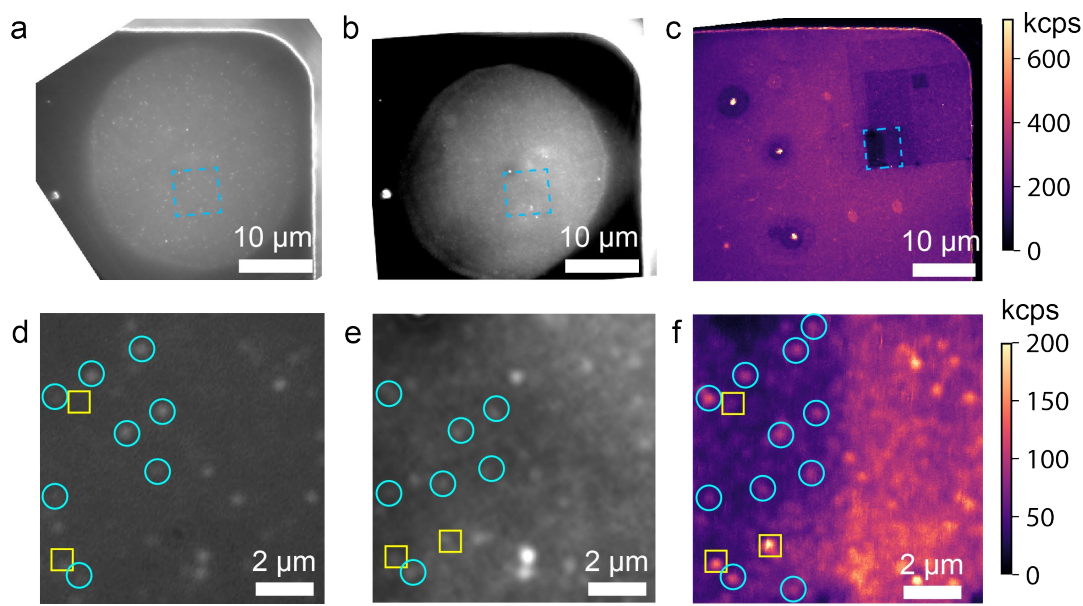

Figure S12: NV⁻ mapping in widefield and confocal images on Sample A. (a) The widefield image of as-prepared Sample A. This image is the same as Figure 4 (a) in the main text but has been rotated to assist alignment. (b)-(c) The same region of Sample A imaged by (b) widefield and (c) confocal microscopy. Compared to (a), Sample A at the time of acquiring (b) and (c) has been incubated with streptavidin and Qdot-525 and has also been subjected to 4 K temperature. These treatments alone or combined has increased the fluorescence background, despite of multiple cycles of solvent cleaning. As a consequence, prior to taking image (c) and subsequent ODMR data, we used high laser intensity to scan and bleach certain regions. The bleached regions are cleanly seen as dark rectangle-shaped regions. (d)-(f) The same 10 µm by 10 µm area, indicated by cyan boxes in (a)-(c), are displayed next to each other. In (f), the emitters highlighted by cyan circles are verified NV⁻ centers with ODMR at 2.87 GHz, while emitters labeled as yellow boxes do not have such ODMR feature. Same symbols are used in (d) and (e) to facilitate visual recognition of the same cluster of emitters via their spatial information (only for the clearly visible ones).

## 5.3 Surface-tethered molecules on diamond membrane

### 5.3.1 Biocompatible diamond surface functionalization

The surface functionalization of the bonded diamond membrane was achieved following the strategy introduced in[25], using 3 % N-[3-(trimethoxysilyl)propyl]ethylenediamine (CAS 1760-24-3, ACROS Organics) in anhydrous acetone for silanization and 0.5 M PEG-SVA solution (dissolved in 100 mM $NaHCO_3$, pH 8.5) that comprises 90 % mPEG-SVA and 10 % biotinPEG-SVA (m.w. 2,000, Laysan Bio) for PEGylation. After functionalization steps, the diamond membrane surface was displaying biotin motifs that served as specific conjugation sites for streptavidin.

### 5.3.2 Flow channel assembly

The coverslip was attached to the bottom of an ibidi sticky-Slide VI 0.4 device (Cat.#: 80608) to form a flow channel for subsequent experiments, and the diamond membrane was facing inwards, as shown in Figure S13. Soaking the device in isopropanol overnight can detach the coverslip for recycling.

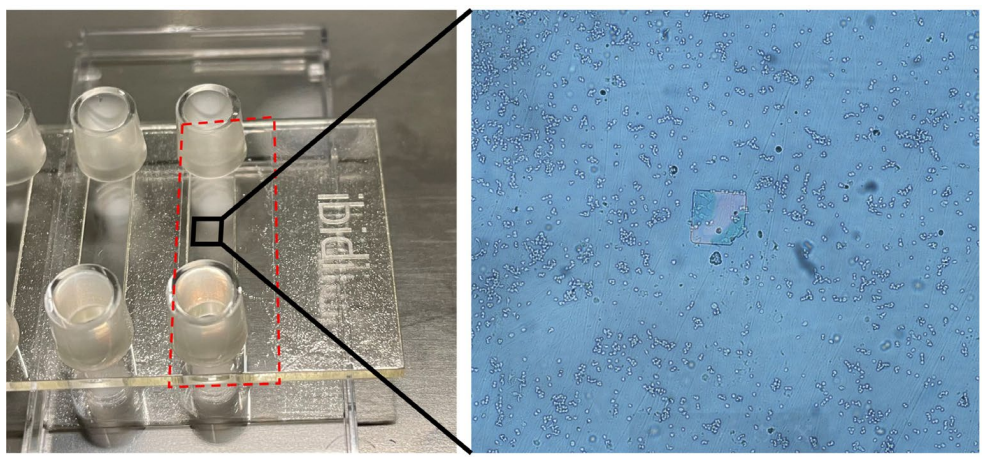

Figure S13: The assembled flow channel device. A fused silica coverslip (red dashed box) with a diamond membrane bonded to the center was attached to a flow channel slide, and RAW cells were incubated, fixed, and stained inside the flow channel, as shown on the right microscopic image.

### 5.3.3 Incubation and imaging

Attachment of target molecules to the functionalized diamond membrane surface was demonstrated using (1) streptavidin labeled with Alexa-488 dye (Invitrogen Cat#S32354), and (2) streptavidin-conjugated Qdot-525 quantum dots (Invitrogen Cat# Q10143MP). First, 60 μL of 5 nm dye-labeled streptavidin solution prepared in 1× phosphate buffered saline (PBS) was introduced to the flow channel and incubated for 5 min at room temperature, before being washed with 1 mL fresh PBS. Widefield fluorescence microscopy images were taken using 488 nm laser (Coherent Sapphire) illumination and 525/50 nm band-pass imaging filter (Chroma). Once the image acquisition was completed, the same field of view was photobleached for 3 min. Second, 50 µL of 10 nM streptavidin-conjugated Qdot-525 was introduced to the flow channel and incubated for 10 min at room temperature before being washed with 1 mL fresh PBS. The same area was imaged, and the fluorescent spots were predominately Qdot-525 as the Alexa-488 conjugated on streptavidin from the previous round of incubation was already photobleached.

### 5.3.4 Simultaneous detection of $NV^-$ centers and Qdot-525

Figure S14 (a) shows the dual-color image simultaneously displaying $NV^-$ centers and Qdot-525. This image was acquired by switching the Hamamatsu W-VIEW GEMINI system to dual-imaging mode. The schematics of this mode is shown in Figure S14 (b). Image pairs from the same field of view were separated by wavelength using a ZT532rdc-UF2 (Chroma) dichroic beam splitter, with a 510/20 nm filter (Chroma ET510/20m) in the short wavelength path and a 594 nm long pass filter (Semrock EdgeBasic LP594) in the long wavelength path. During a single exposure which lasted 100 s, 488 nm laser was switched on only for the first 10 s while the 532 nm laser was on for the entire duration to help balance the overall intensities of the two sub-images.

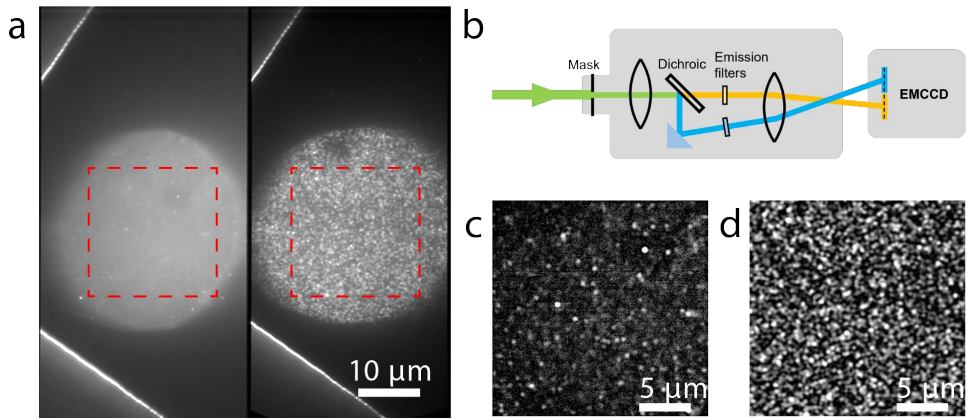

Figure S14: Simultaneous detection of both $NV^-$ and Qdot-525 on Sample A. (a) A single-exposure image acquired on a 1024×1024 pixel EMCCD, showing two sub-images of the same area differentiated by emission wavelengths. The left sub-image shows $NV^-$ and the right one shows Qd-525. (b) Schematic illustration of Hamamatsu W-VIEW GEMINI system. (c)-(d) The zoomed-in view of the regions in (a) indicated by red boxes, respectively. The images were processed in ImageJ to subtract background and improve contrast for visualization. For real-world applications, careful system calibration using fluorescent beads is required for the accurate alignment of the two sub-images to remove small yet systematic distortion. This framework enables the selection of $NV^-$-target molecule pairs that are spatially closed to each other for subsequent quantum sensing experiments.

## 5.4 RAW cell experiments

We duplicated a new flow channel device for cell culturing. The flow channel was washed with culturing medium (DMEM with 10 % FBS) and $1 \times 10^5$ RAW cells were delivered and incubated at 37 °C and 5 % $CO_2$ overnight. The channel was washed with fresh medium and the cells were fixed by incubating with 3 % of paraformaldehyde for 15 min. The channel was washed again and incubated for 60 min with 1:100 of 0.5 mg mL$^{-1}$ anti-TLR2 monoclonal antibody (Invitrogen) labeled with Alexa-488 following a standard protocol.[26] Finally, the channel was washed 3 times with PBS for subsequent imaging. Widefield images under 488 nm laser excitation was acquired by EMCCD with 1 s exposure time and 300 gain. For each region, the first image was acquired at a very steep TIRF angle, followed by a second image that was acquired under epi-illumination.

## 5.5 Bacteria experiments

*Escherichia coli* (*E. coli*, BL21 strain) bacteria that overexpress green fluorescent protein (GFP) were suspended in PBS and introduced to another flow channel. Initially, very few bacteria cells were seen to settle on the diamond membrane surface using fluorescence microscopy. After approximately 6 h of sediment, a much higher density of cells was observed on the diamond membrane surface and images/videos were taken with 488 nm laser illumination under this condition. See Supplementary Video S1 as an example.

## 5.6 TIRF condition and evanescent field profile

The depth resolution of TIRF microscopy originates from the rapid evanescent field decay of light incident at angles greater than the critical angle. The critical angle of our system, referenced in glass, is $\theta_c = sin^{-1}(n_{water}/n_{glass}) \approx 66°$. We used finite difference time domain simulations (Lumerical, FDTD) to calculate the electric intensity profile across the diamond membrane stack. Figure S15 (a) is an illustrative plot of the normalized electric field intensity for p-polarized light, incident at an angle of 70° on a 160 nm-thick diamond membrane; the field decays exponentially into the water layer. Figure S15 (b) is a plot of the extracted decay constant in the water layer computed from both FDTD and the analytic expression $\frac{\lambda_0}{4\pi}(n_1^2 \sin^2(\theta) - n_2^2)^{-\frac{1}{2}}$ for incident angles greater than $\theta_c$. Such large incident angles are accessible in our setup due to the 1.5 N.A. of the oil-immersion objective. From this analysis, we expect our diamond-membrane TIRF configuration to have depth resolution ≤ 100 nm.

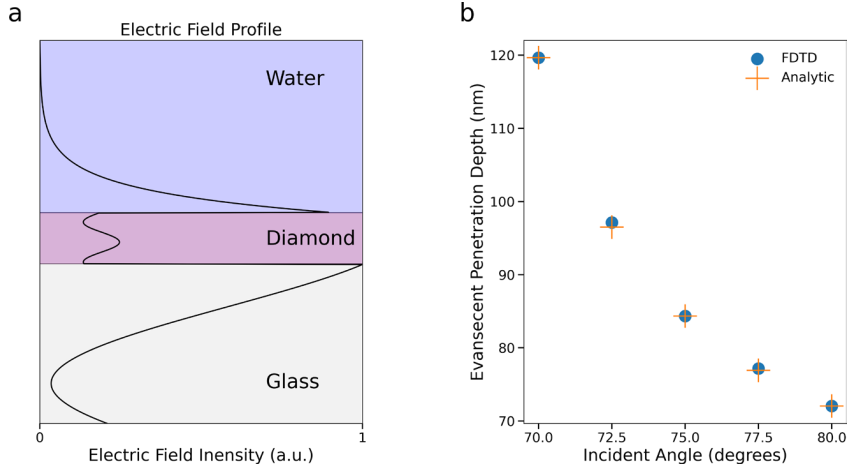

Figure S15: Calculated intensity profile in diamond-membrane TIRF microscopy. (a) FDTD-calculated electric field intensity profile across a 160 nm-thick diamond membrane between a glass substrate and water. Here, the light is p-polarized and incident at an angle of 70°, as referenced in the glass substrate. (b) Intensity decay constant in the water layer calculated using FDTD (blue circle) and analytically (orange cross) for angles greater than the critical angle.

## 5.7 Spin coherence of NV$^-$ centers in direct-bonded membranes

The spin coherence of NV$^-$ centers is critical to their performance as nanoscale quantum sensors. In a previous study[1], we have experimentally measured technology-compatible room temperature spin coherence of individual NV$^-$ centers in diamond membranes, reflecting the pristine crystal quality and atomically smooth surfaces of the membrane. In this section, we investigate the impact of the direct bonding process on NV$^-$ coherence to determine if the processes introduce additional decoherence processes. The device consists of a 150 nm-thick diamond membrane bonded to a thermal oxide wafer. The membrane contains implanted and naturally formed $^{14}$NV$^-$ centers, as well as naturally formed $^{15}$NV$^-$ centers. The existence of background $^{15}$NV$^-$ originates from the diamond growth chamber which is regularly used for $\delta$-doping of [$^{15}$N]. Measurements are performed at room temperature, with the microwave drive introduced via a coplanar waveguide patterned onto a glass coverslip, as shown in Figure S16 (g). Two NV$^-$ centers were randomly chosen and characterized, one with an intrinsic [$^{15}$N] while the other with [$^{14}$N]. Both centers were identified via the ODMR features shown in Figure S16 (a)-(b). The Ramsey measurement shown in Figure S16 (c)-(d) shows $T_2^*$ of 92(16) μs and 66(9) μs, respectively. The Fourier transform of $T_2^*$ measurements with a much finer scan (to avoid undersampling) are shown in Figure S16 (e)-(f). We note that the $^{15}$NV$^-$ (Figure S16 (e)) is coupled to a nearby $^{13}$C nuclear spin, which results in an oscillation pattern in the spin echo measurement, as shown in Figure 2 (g) of the main text with a Hahn echo $T_2$ of 632(21) μs ($T_2$ fitting superscript $n = 1.35 \pm 0.04$). The fitting details of Hahn echo $T_2$ is described below.

We first analyze the interaction term from the weakly coupled $^{13}$C. In the rotational frame of the NV$^-$ electron spin the Hamiltonian is given by:

$$H = \omega_n I_z + S_z(AI_z + BI_x) = (Am_s - \omega_n)I_z + Bm_s I_x \quad (1)$$

Where $m_s \in \{0, 1\}$ is determined by the electron spin state. Diagonalizing the Hamiltonian gives the resonance frequency $K = \sqrt{(Am_s - \omega_n)^2 + (Bm_s)^2}$, such that $K_+ = \sqrt{(A - \omega_n)^2 + (B)^2}$ and $K_- = \omega_n$. The fitting function can thus be simplified to the following form:

$$f(\tau) = Ae^{\left(-\frac{\tau}{T_2}\right)^n}\left[1 + 2\left(\frac{B\omega_n}{K_+ K_-}\right)^2 \sin^2\left(\frac{2\pi K_+ \tau}{4} + \phi_0\right)\sin^2\left(\frac{2\pi K_- \tau}{4} + \phi_1\right)\right]$$

$$= Ae^{\left(-\frac{\tau}{T_2}\right)^n}\left[1 + 2\left(\frac{B}{K_+}\right)^2 \sin^2\left(\frac{2\pi K_+ \tau}{4} + \phi_0\right)\sin^2\left(\frac{2\pi K_- \tau}{4} + \phi_1\right)\right] \quad (2)$$

Notice that $\tau$ in our case is defined to be the time from the beginning of the first $\frac{\pi}{2}$ pulse to the end of the last $\frac{\pi}{2}$, as such there is an additional factor of 2 compared to the formula from the original paper.[27]

The modulation frequencies—97.32(5) kHz and 14.82(5) kHz—correspond respectively to the free precession of $^{13}$C at $\sim$ 91 Gauss (the magnetic field used for spin measurements) and the combination of free precession and coherent coupling strength of the $^{13}$C to the $^{15}$NV$^-$. Fitting our spin echo data returns $B = 13.14(14)$ kHz. From our derivation above, we obtain $A = \omega_n \pm \sqrt{K_+^2 - B^2}$, which is 90.493 kHz or 104.176 kHz. We can make a rough estimate of the separation between the $^{13}$C and our NV center by ignoring the Fermi contact interaction. This can be done as follows:

$$A = \frac{\mu_0 \gamma_e \gamma_n \hbar}{4\pi r^3}(3\cos^2(\theta) - 1) \quad (3)$$

$$B = \frac{\mu_0 \gamma_e \gamma_n \hbar}{4\pi r^3} 3\cos(\theta)\sin(\theta)$$

Solving for $\theta$ and $r$ gives:[28]

$$\theta = \arctan\left(\frac{1}{2}\left(-3\frac{A}{B} + \sqrt{9\frac{A^2}{B^2} + 8}\right)\right)$$

$$r = \left(\frac{\mu_0 \gamma_e \gamma_n \hbar (3\cos^2(\theta) - 1)}{4\pi A}\right)^{\frac{1}{3}} \quad (4)$$

We then have $r = 1.40$ nm with $\theta = 5.5°$ or $r = 1.33$ nm with $\theta = 4.8°$ corresponding to $A = 90.493$ or 104.176 kHz.

In Figure 2 (g) of the main text, a well-aligned magnetic field along the NV$^-$ axis suppresses the $^{15}$N modulation on the spin echo signal. In contrast, in Figure S16 (h) $^{14}$NV$^-$'s large quadrupole interaction at low magnetic fields suppresses modulation regardless of magnetic field alignment[29]. Therefore, it returns a simple decay without modulation and a corresponding Hahn echo $T_2$ of 278(13) μs ($T_2$ fitting superscript $n = 1.30 \pm 0.06$). We note that the coherence time is comparable to the NV$^-$ centers' value measured at the suspended region[1], indicating negligible added noise from the direct bonding process. The distribution of $T_2$ is affected by the depth of NV$^-$ centers as discussed in Ref.[30], and this effect becomes more apparent in our diamond membrane system where two surfaces are present. In the future, a more rigorous characterization regarding optical and spin coherence of shallow NV$^-$ centers in diamond membranes will be conducted, which is beyond the scope of the current work.

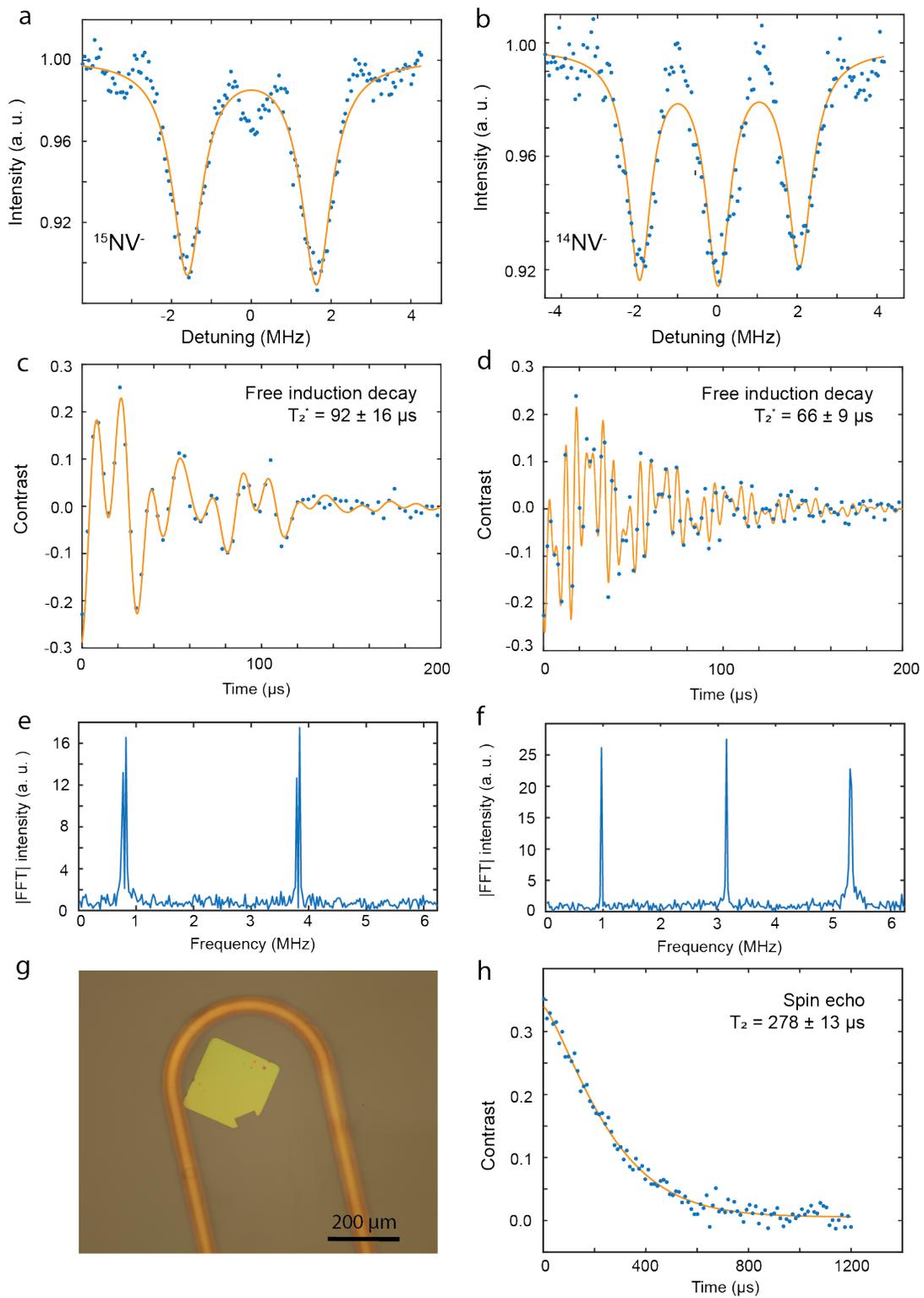

Figure S16: Typical spin coherence of NV centers in direct-bonded membranes at room temperature. (a), (c), (e) were measured on a single $^{15}$NV$^-$ while (b), (d), (f), (h) were measured on a single $^{14}$NV$^-$. (a)-(b): the ODMR spectra. We simplified the fitting using multi-peak Lorentzian to retrieve the ODMR frequency and linewidth. A more careful fitting with multiple Rabi formula curves could lead to a more precise fit. (c)-(d): Free induction decay (Ramsey measurement) of the NV$^-$s. Both NV$^-$s show notably long $T_2^*$ times due to the isotopic purification during overgrowth. (e)-(f): Fourier transformations of finely sampled Ramsey measurements show the Larmor precession frequencies coming from in e) the intrinsic $^{15}$N with additional splitting from a distant $^{13}$C and in (f) only the intrinsic $^{14}$N. (g) A bright-field microscope image of the Ω-shape coplanar waveguide used for microwave signal delivery to the diamond membrane. (h) Hahn echo measurements of the $^{14}$NV$^-$ showing a long $T_2$ value with a simple decay.

# 6 Additional XPS figures

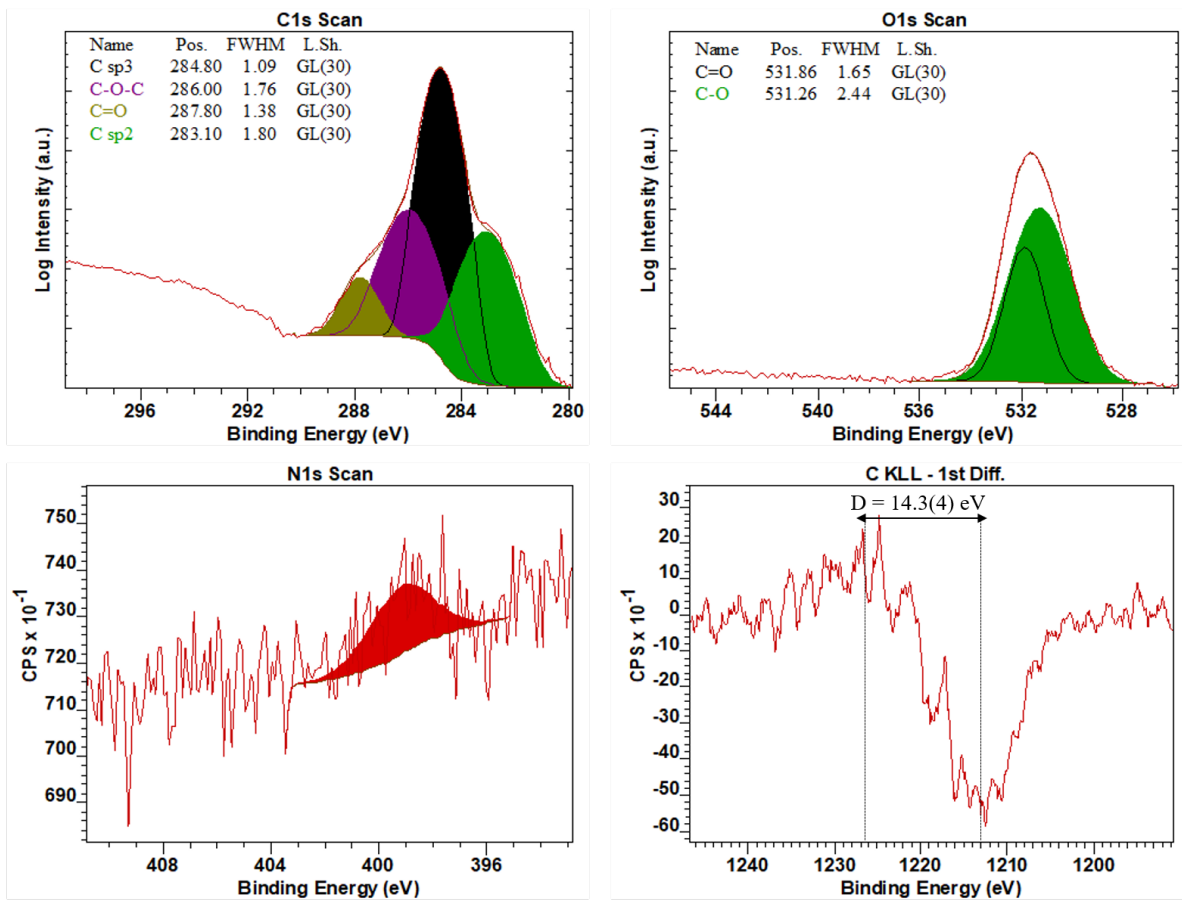

Figure S17: XPS characterization of the processed diamond substrate, showing the C 1s and O 1s deconvolutions with component labels and fitting parameters. $sp^2$ content quantification is achieved via D parameter linear extrapolation of the C KLL first order derivative[12,13]. N 1s peak contamination also showed as samples were exposed to atmosphere. Some of the peak fittings are presented in log scale to help discern minority components.

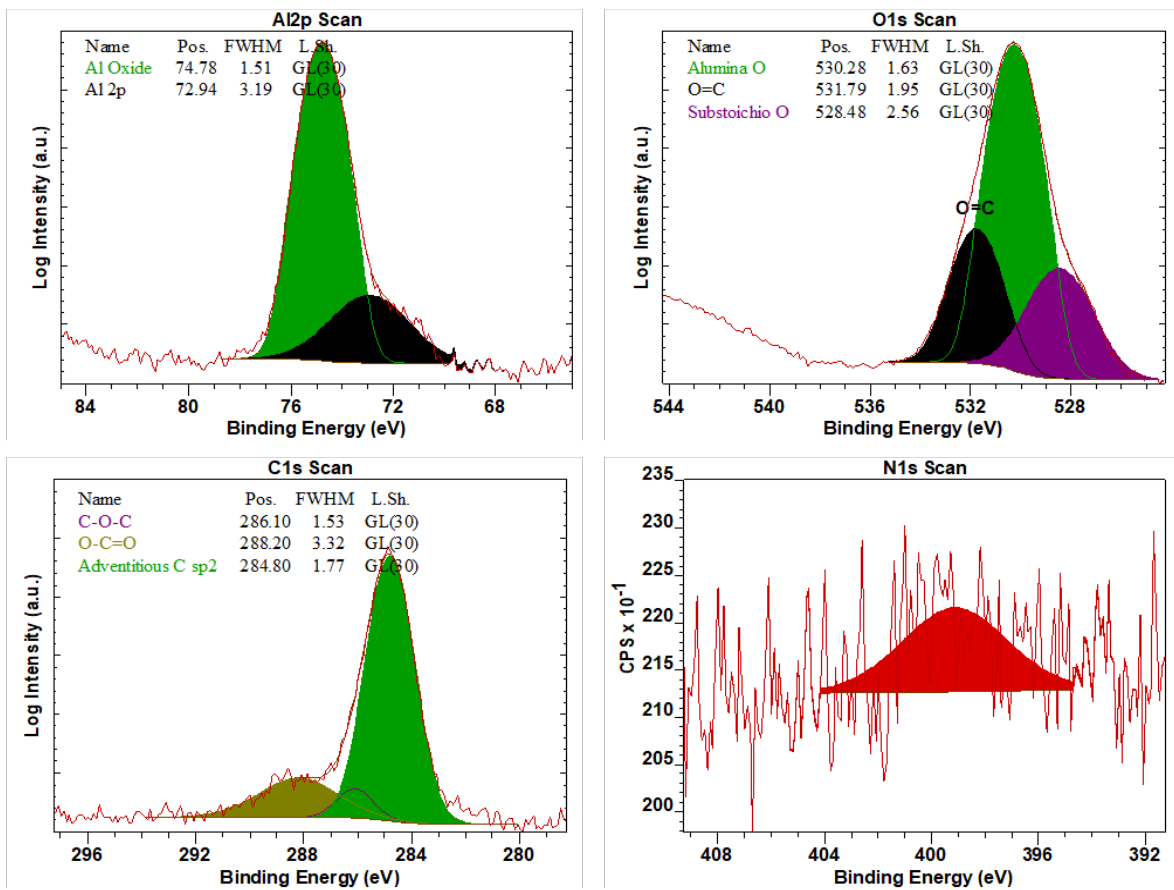

Figure S18: XPS characterization of the unprocessed sapphire substrate, showing the Al 2p, O 1s, and C 1s deconvolutions with component labels and fitting parameters. N 1s peak contamination also showed as samples were exposed to atmosphere. Some of the peak fittings are presented in log scale to help discern minority components.

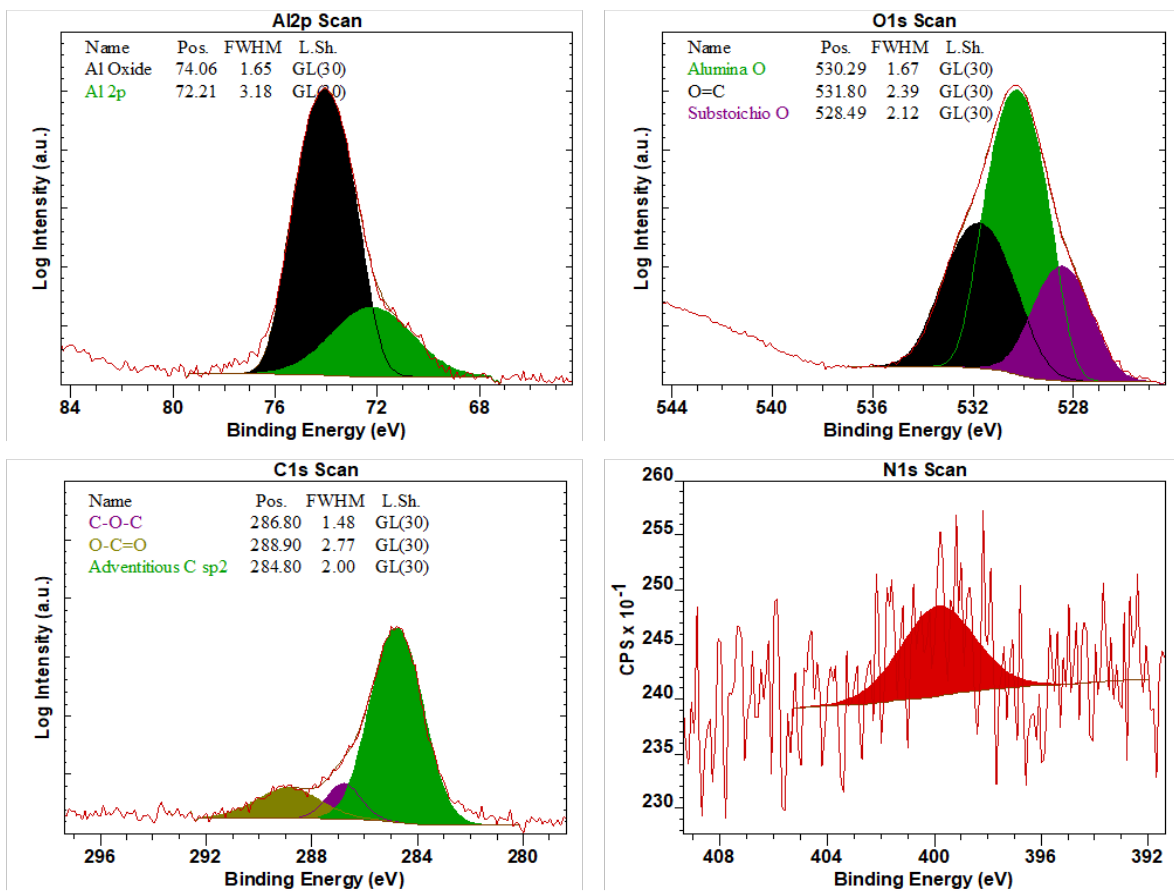

Figure S19: XPS characterization of the processed sapphire substrate, showing the Al 2p, O 1s, and C 1s deconvolutions with component labels and fitting parameters. N 1s peak contamination also showed as samples were exposed to atmosphere. Some of the peak fittings are presented in log scale to help discern minority components.

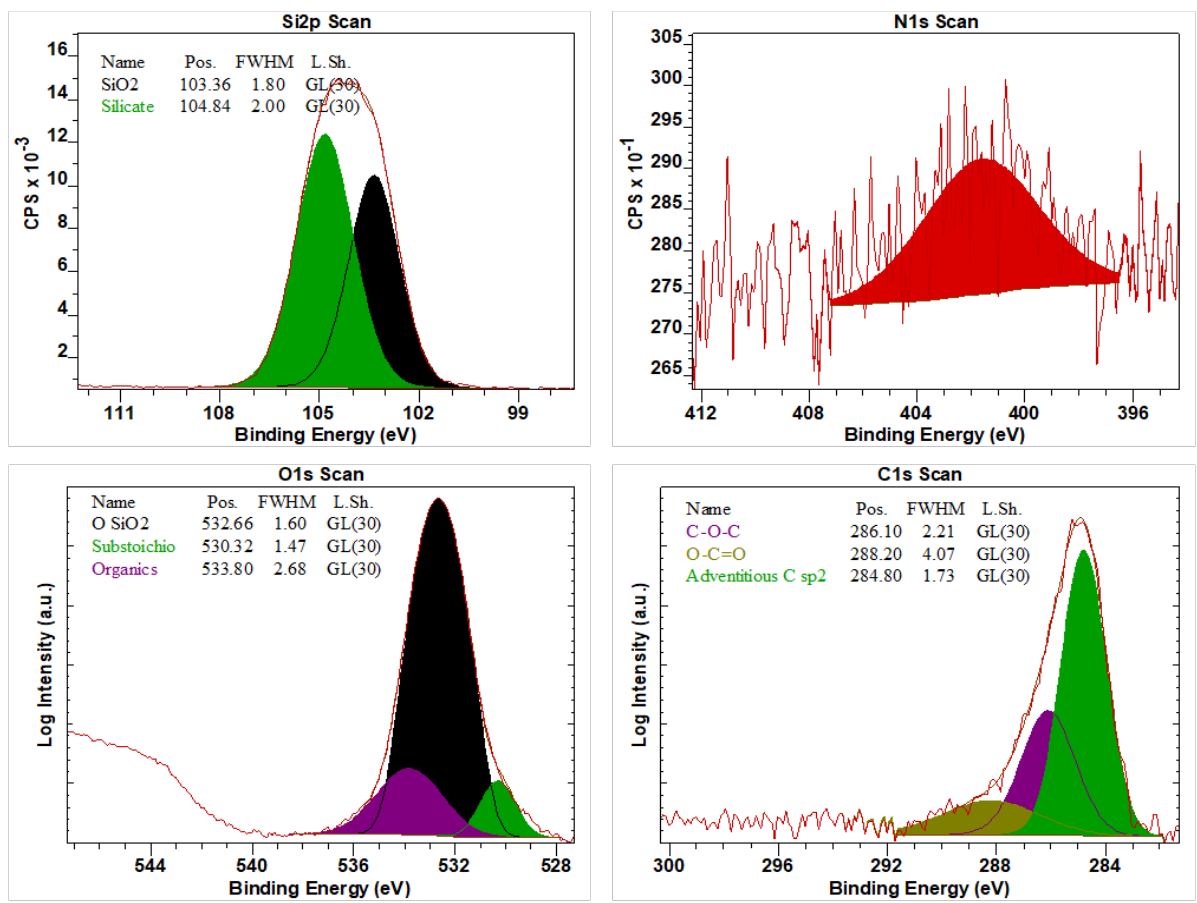

Figure S20: XPS characterization of the unprocessed sapphire substrate, showing the Si 2p, O 1s, and C 1s deconvolutions with component labels and fitting parameters. N 1s peak contamination also showed as samples were exposed to atmosphere. Some of the peak fittings are presented in log scale to help discern minority components.

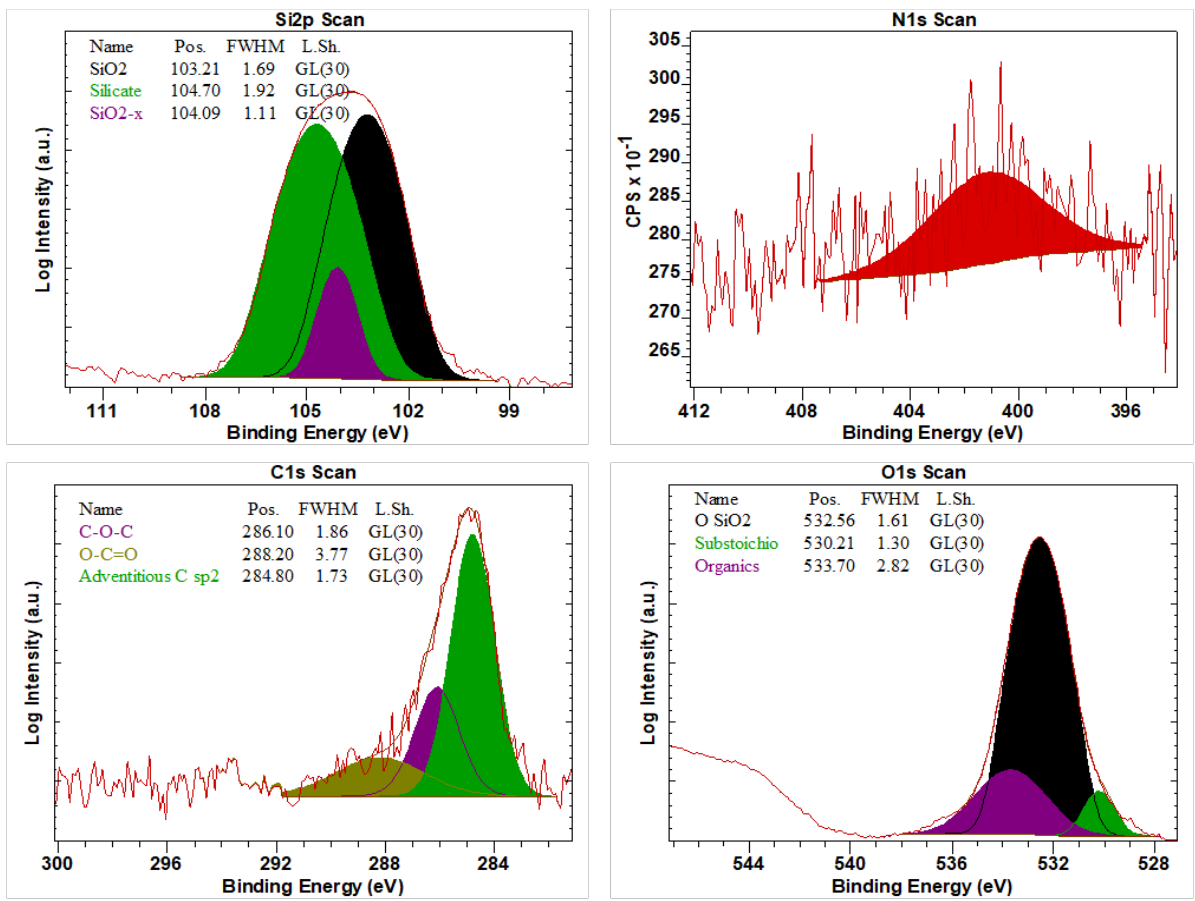

Figure S21: XPS characterization of the processed sapphire substrate, showing the Si 2p, O 1s, and C 1s deconvolutions with component labels and fitting parameters. N 1s peak contamination also showed as samples were exposed to atmosphere. Some of the peak fittings are presented in log scale to help discern minority components.

# 7 Supplementary videos (supplied as separate files)

Supplementary Video S1: The movie shows a series of widefield fluorescence images of *E. coli* bacteria overexpressing GFP that sedimented on the diamond membrane surface inside a flow channel. The movie is displayed in real time. The field-of-view corresponds to $73 \times 73$ μm$^2$. Edges of diamond membrane are also visible.